\documentclass[manuscript,endfloat,noblind]{geophysics}
\usepackage{graphicx}
\usepackage{subfigure}

\begin{document}

\title{The role of the background velocity model for the Marchenko focusing of reflected and refracted waves}
\renewcommand{\thefootnote}{\fnsymbol{footnote}} 

\ms{submitted to \textit{Geophysical Journal International}} 
\author{Mert Sinan Recep Kiraz$^1$\thanks{Corresponding author}, Roel Snieder$^1$, and Kees Wapenaar$^2$ \\
  $^1$ Center for Wave Phenomena, Colorado School of Mines, Golden CO 80401, USA \\ E-mail: mertkiraz@gmail.com; rsnieder@mines.edu \\
  $^2$ Department of Geoscience and Engineering, Delft University of Technology, P.O. Box 5048, Delft, GA 2600, The Netherlands \\ E-mail: c.p.a.wapenaar@tudelft.nl
  }

\righthead{Computational seismology, Seismic interferometry, Wave propagation}

\maketitle

\clearpage
\newpage

\begin{abstract}
Marchenko algorithms retrieve the wavefields excited by virtual sources in the subsurface, these are the Green's functions consisting of the primary and multiple reflected waves. The requirements for these algorithms are the same as for conventional imaging algorithms; they need an estimate of the velocity model and the recorded reflected waves. We investigate the dependence of the retrieved Green's functions using the Marchenko equation on the background velocity model and address the question: ``How well do we need to know the velocity model for accurate Marchenko focusing?". We present three different background velocity models and compare the Green's functions retrieved using these models. We show that these retrieved Green's functions using the Marchenko equation give correlation coefficients with the exact Green's function larger than 90\%  on average except near the edges of the receiver aperture. We also examine the presence of refracted waves in the retrieved Green's function. We show with a numerical example that the Marchenko focusing algorithm produces refracted waves only if the initial velocity model used for the iterative scheme is sufficiently detailed to model the refracted waves.
\end{abstract}

\section{Introduction}

The inverse scattering community utilized the Marchenko equation to relate scattered data to the scattering potential to determine the medium properties \cite[]{Marchenko55Construction,Gelfand55IS,Agranovich63,newton1d,burridge,ChadanSabatier,Gladwell93,ColtonKress}. The connection between wavefield focusing and the Marchenko equation was made by \cite{Rose2001, rosebook} so that the wavefield focusing at a location in an unknown medium can be achieved once the Marchenko equation is solved. \cite{BrogginiandSnieder} connect the Marchenko equation and seismic interferometry \cite[]{WeaverandLobkis,Derode03How,Wapenaaretal2005,Curtis06Noise,Snieder13Annrev} and show that one can retrieve the Green's function between any point in the subsurface and points on the acquisition surface without a physical receiver at the virtual source location and with one-sided illumination. \cite{wapenaar3d} discuss the three-dimensional Green's function retrieval, and present an example of the two-dimensional Green's function retrieval. 

A thorough description of the Marchenko redatuming and imaging method and its numerical implementation is given by \cite{Wapenaaretal2014}, \cite{marchenkoleading}, \cite{thobeckimpl}, and \cite{lomasandcurtis}. Marchenko methods have been widely used for various applications such as internal multiple elimination \cite[]{meles2,meles1,thorbecke21}, elastic wave applications \cite[]{dacosta1,dacosta2,wapenaarelastic}, subsurface imaging and for comparisons with the conventional imaging results (e.g., reverse time migration) \cite[]{behuraetal,Wapenaaretal2014,ravasietal,jiaetal}. Additionally, various field data set applications of the Marchenko method have been performed such as imaging of a North Sea field data set \cite[]{ravasietal}, target-oriented subsalt imaging of the Gulf of Mexico data set \cite[]{jiaetal}, time-lapse monitoring of the Frio carbon sequestration data set \cite[]{kirazandnowack}, multiple suppression on an Arabian Gulf field data set \cite[]{Staring2021GEO}, and an offshore Brazil data set for imaging a reservoir under basalt \cite[]{Jia21Subsalt}. With growing interest in machine learning applications in seismology, a convolutional neural network-based 1D wavefield focusing is also proposed by \cite{kirazcnn}.

Recent studies have aimed to address the limitations of the up/down separation of the Marchenko equation. Using the data collected on a closed received array, \cite{kirazetal,kiraz2021} retrieve the full-field (e.g., without component separation) Green's function, and show that the full-field Marchenko focusing provides better focusing results in the subsurface than achievable using only the direct waves. \cite{leonivan} and \cite{wapenaar2021greens} show alternative methods where one can retrieve the Green's function using single-sided acquisition data without up/down decomposition.

In this paper, we use the one-sided Marchenko focusing to retrieve the Green's function at an arbitrary depth location using different subsurface models with variable velocity and variable density profiles. In Section 2, we describe the Marchenko focusing algorithm and show that it requires only two inputs; surface-recorded data and the initial estimate of the velocity model, which are the same inputs as for conventional imaging algorithms. In Section 3, we provide a visual tour to describe the iterative Marchenko focusing algorithm. In Section 4, we investigate the background velocity model dependence of the Marchenko method and show the accuracy of the retrieved Green's function by presenting correlation coefficients (CCs) between the retrieved and numerically modeled Green's functions. Lastly, in Section 4, we use the Marmousi model to investigate the presence of the refracted waves in the Marchenko focusing, and show that the presence of the refracted waves depends only on the initial estimate of the velocity model.

\section{Methodology}\label{sec:2}

We use the Marchenko algorithm proposed by \cite{wapenaar3d} which builds on earlier work of \cite{Rose2001}, \cite{rosebook}, and \cite{BrogginiandSnieder}. We denote spatial coordinates as $\textbf{x}=(x,z)$, and the receiver coordinates as $\textbf{x}_{R}=(x_{R},z_{R})$. {The receivers are located at the transparent acquisition surface at $z=0$, and the multiples caused by free surface (e.g., air-water interface) are excluded}. 

We relate the ingoing wave, $U_{k}^{in}$, to the outgoing wave, $U_{k}^{out}$, at iteration $k$ as
\begin{equation}
    U^{out}_{k}(\textbf{x}_{R},t)=\int R(\textbf{x}_{R},\textbf{x},t) \ast U^{in}_{k}(\textbf{x},t) dx\;,
    \label{eq:maintime}
\end{equation}
where $R$ corresponds to the reflection response of the medium and the asterisk denotes temporal convolution. The iterative scheme starts with modeling the direct wave, and we denote the arrival time of the direct wave from the virtual source location, $\textbf{x}_s$, for which we aim to retrieve the Green's function, to the receivers at the surface as $t_d$. Following \citet{BrogginiandSnieder}, we design an iterative scheme so that at $t=0$, the wavefield becomes a delta function at the pre-defined focal (or virtual source) location. We start the iterative algorithm by defining the ingoing wavefield at $z=0$ for iteration $k$ as
\begin{equation}
    U^{in}_{k}(\textbf{x},t)=U^{in}_{0}(\textbf{x},t)-\Theta(\textbf{x},t)U^{out}_{k-1}(\textbf{x},-t)\;,
    \label{eq:highit}
\end{equation}
 where $\Theta(\textbf{x},t)$ defines a window function where $\Theta(\textbf{x},t)=1$ when $-t_{d}^{\epsilon}(\textbf{x}) < t < t_{d}^{\epsilon}(\textbf{x})$ and otherwise $\Theta(\textbf{x},t)=0$ with $t_{d}^{\epsilon}=t_{d}-\epsilon$ where we introduce $\epsilon$ as a small positive constant to exclude the direct wave at $t_d$. After the convergence is achieved (i.e., $U^{out}_{k}=U^{out}_{k-1}$) we can drop the iteration number, and insert equation~(\ref{eq:highit}) into equation~(\ref{eq:maintime}), and obtain
\begin{equation}
        U^{out}(\textbf{x}_{R},t)=\int R(\textbf{x}_{R},\textbf{x},t) \ast U^{in}_{0}(\textbf{x},t)\;dx -\int R(\textbf{x}_{R},\textbf{x},t)\ast \Theta(\textbf{x},t)U^{out}(\textbf{x},-t)\;dx\;.
        \label{eq:comp}
\end{equation}

Once the convergence is achieved, we denote the recorded data at the receivers as $U_{total}(\textbf{x},t)=U^{in}(\textbf{x},t)+U^{out}(\textbf{x},t)$ which consists of the superposition of the ingoing and outgoing wavefield. We then define $p(\textbf{x},t)$ as the total wavefield that is associated with $U_{total}(\textbf{x},t)$. We obtain the homogeneous Green’s function  ($G_{h}(\textbf{x},\textbf{x}_{s},t)=G(\textbf{x},\textbf{x}_{s},t)-G_(\textbf{x},\textbf{x}_{s},-t)$) \cite[]{Oristaglio89}, for the virtual source location $\textbf{x}_{s}$ and the receiver location $\textbf{x}$ as
\begin{equation}
    G_{h}(\textbf{x},\textbf{x}_{s},t) = p(\textbf{x},t) - p(\textbf{x},-t)\;.
    \label{eq:orishom}
\end{equation}
Equation (\ref{eq:orishom}) satisfies the homogeneous wave equation and, hence, retrieves the Green’s function for times $t>0$ for the virtual source location $\textbf{x}_s$ \cite[]{Oristaglio89,wapenaar3d,kiraz2021}. The iterative scheme we use follows the algorithm presented by \cite{wapenaar3d}, and equations (\ref{eq:maintime}), (\ref{eq:highit}), and (\ref{eq:orishom}) are given in \cite{wapenaar3d} in equations (13), (12), and (14), respectively.

\begin{figure}
  \centering
  \subfigure[]{\includegraphics[width=0.6\textwidth]{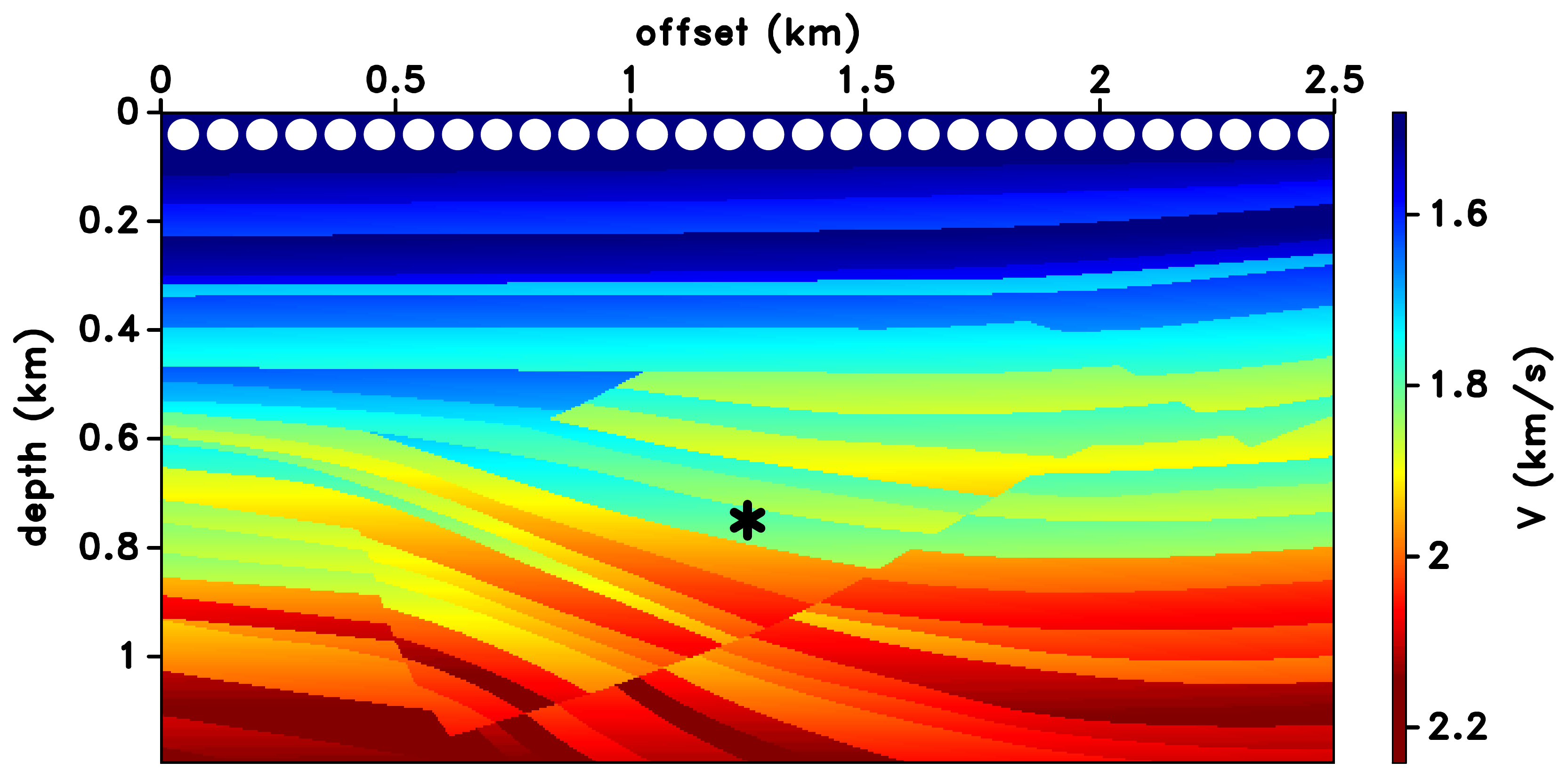}\label{fig:vel1}}
  \subfigure[]{\includegraphics[width=0.6\textwidth]{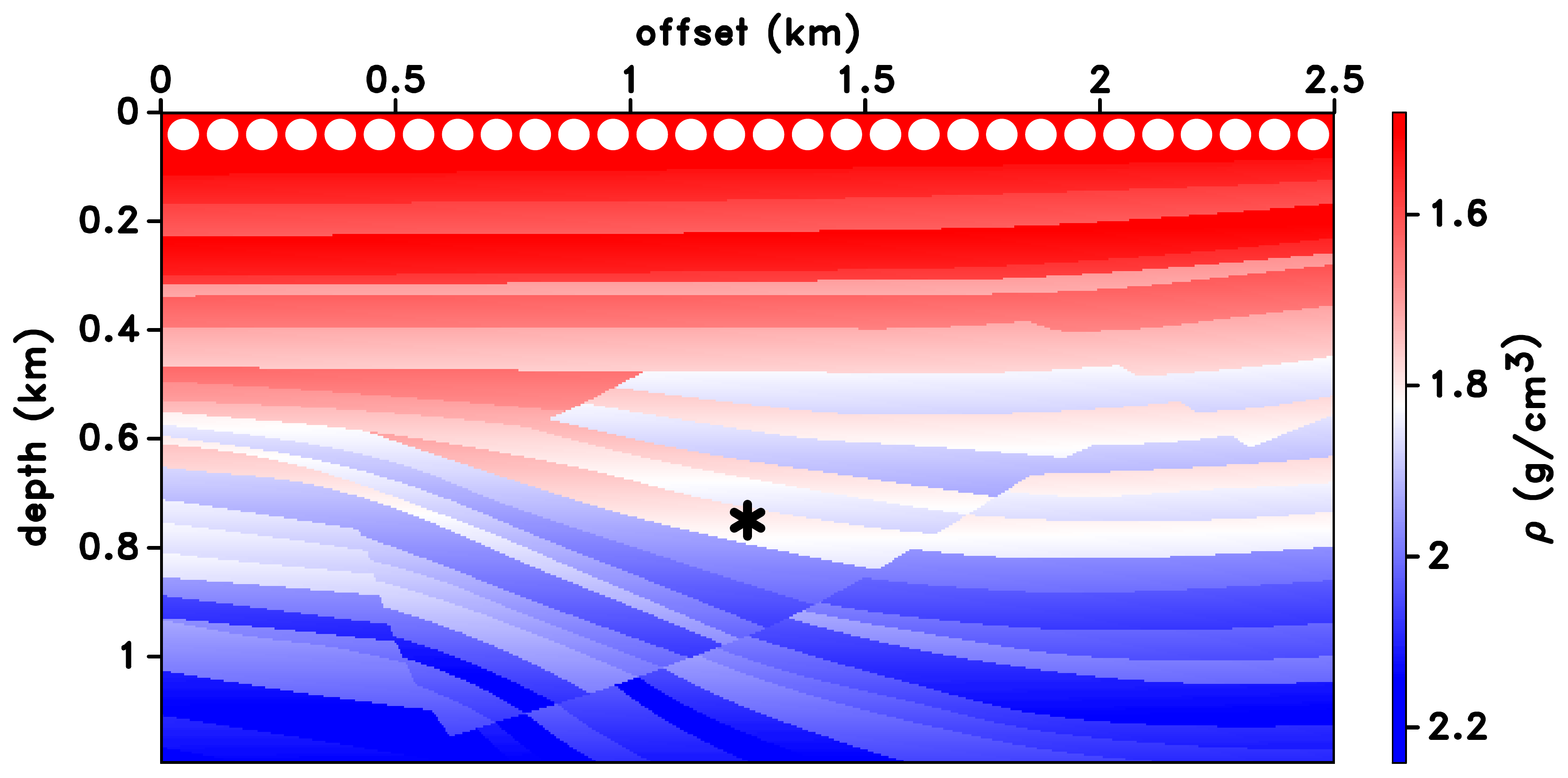}\label{fig:den1}}
  \caption{(a) Velocity model and (b) density model of the synthetic example which are extracted from the Sigsbee model. The black asterisk shows the virtual source location and the white dots at the top indicate every 30th source/receiver location.}
  \label{fig:example1}
\end{figure}

\section{Visualizing the iterative scheme}
In this section, we present a numerical example to aid the visual understanding of the iterative scheme described in Section 2. Figure \ref{fig:example1} shows the subsurface model and source and receiver geometry of our first numerical experiment. Figures \ref{fig:vel1} and \ref{fig:den1} show the variable velocity and density models used for the numerical example, respectively, which are extracted from the Sigsbee model \cite[]{Paffenholz} with 2.5 km horizontal and 1.2 km vertical extent. The virtual source location is at $x_s$ = 1.25 km and $z_s$ = 0.75 km in depth which is shown with the black asterisk in Figure \ref{fig:example1}. The white dots located at the surface of the models in Figure \ref{fig:example1} represent every 30th receiver location. {We use point impulsive sources and record pressure at the receivers, and exclude the presence of free surface}. {During the iterative Marchenko scheme, we use the normal derivative of the pressure field to send the wavefield into the medium from the receiver array} \cite[]{kiraz2021}.

\begin{figure}
  \centering
  \includegraphics[width=0.6\textwidth]{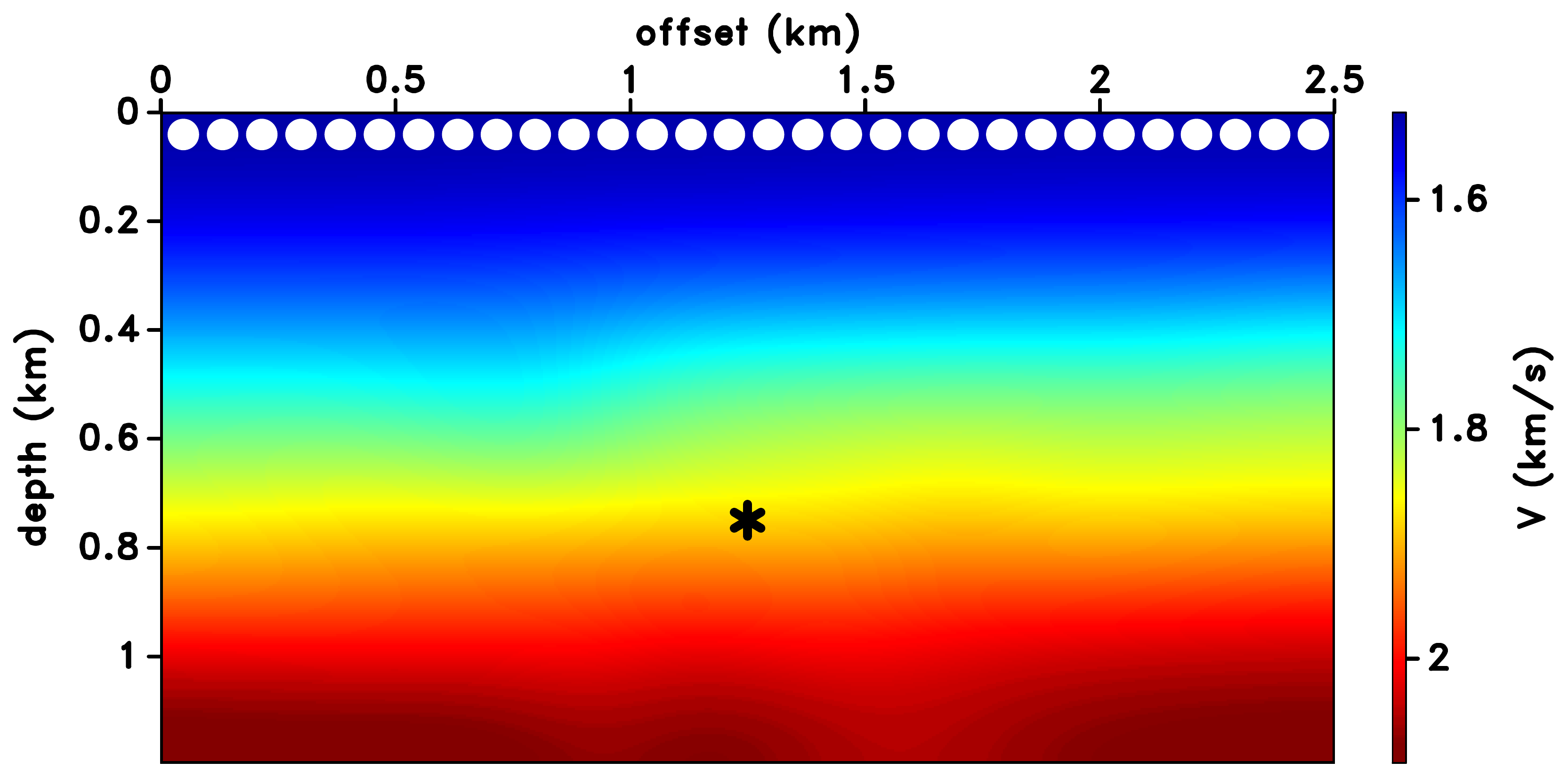}
  \caption{Smoothed version of the velocity model used for the iterative algorithm. The black asterisk shows the virtual source location and the white dots at the top indicate every 30th source/receiver location.}
  \label{fig:velsmooth1}
\end{figure}


\begin{figure}
  \centering
  \subfigure[]{\includegraphics[width=0.4\textwidth]{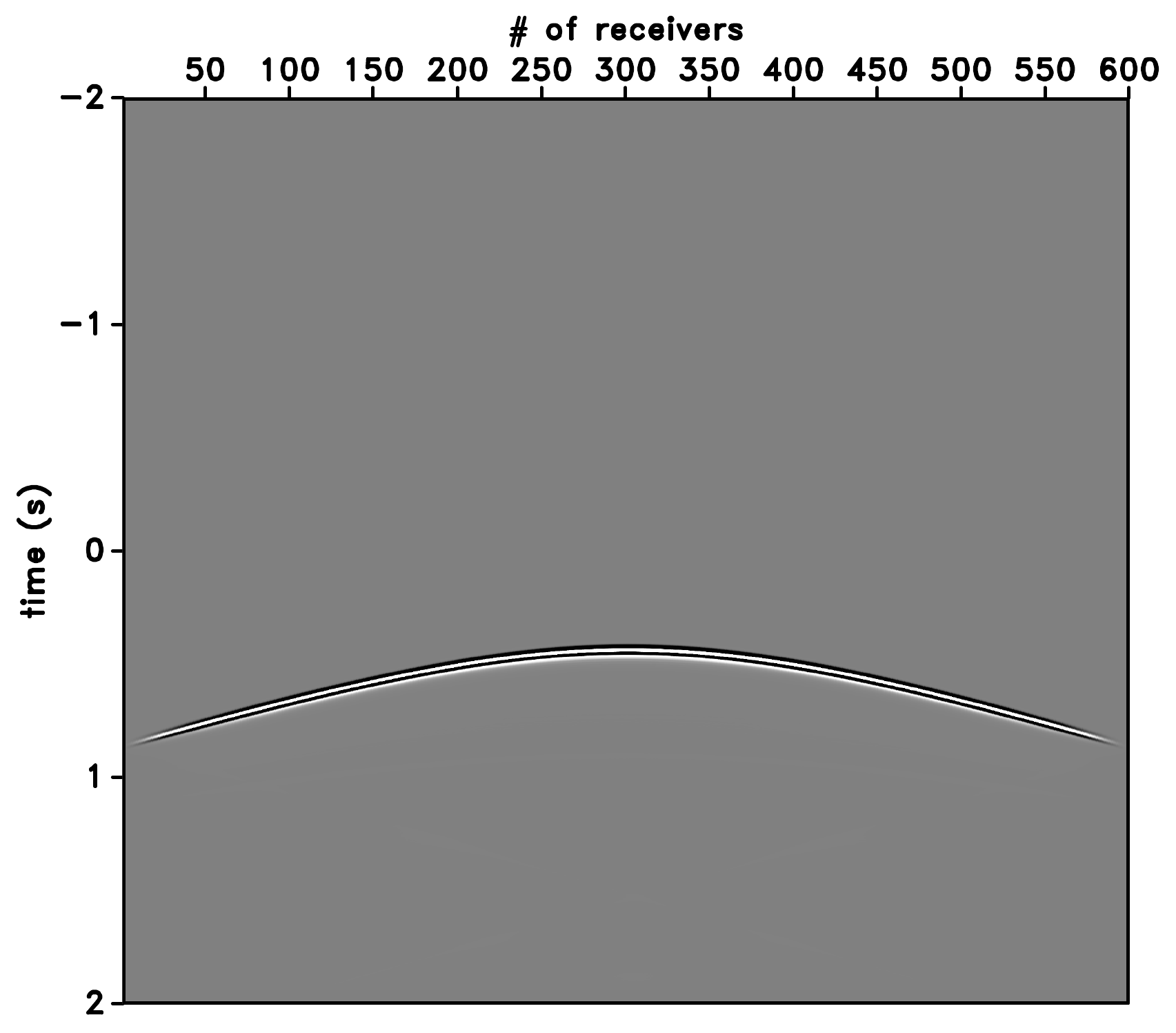}\label{fig:it1}}
  \subfigure[]{\includegraphics[width=0.4\textwidth]{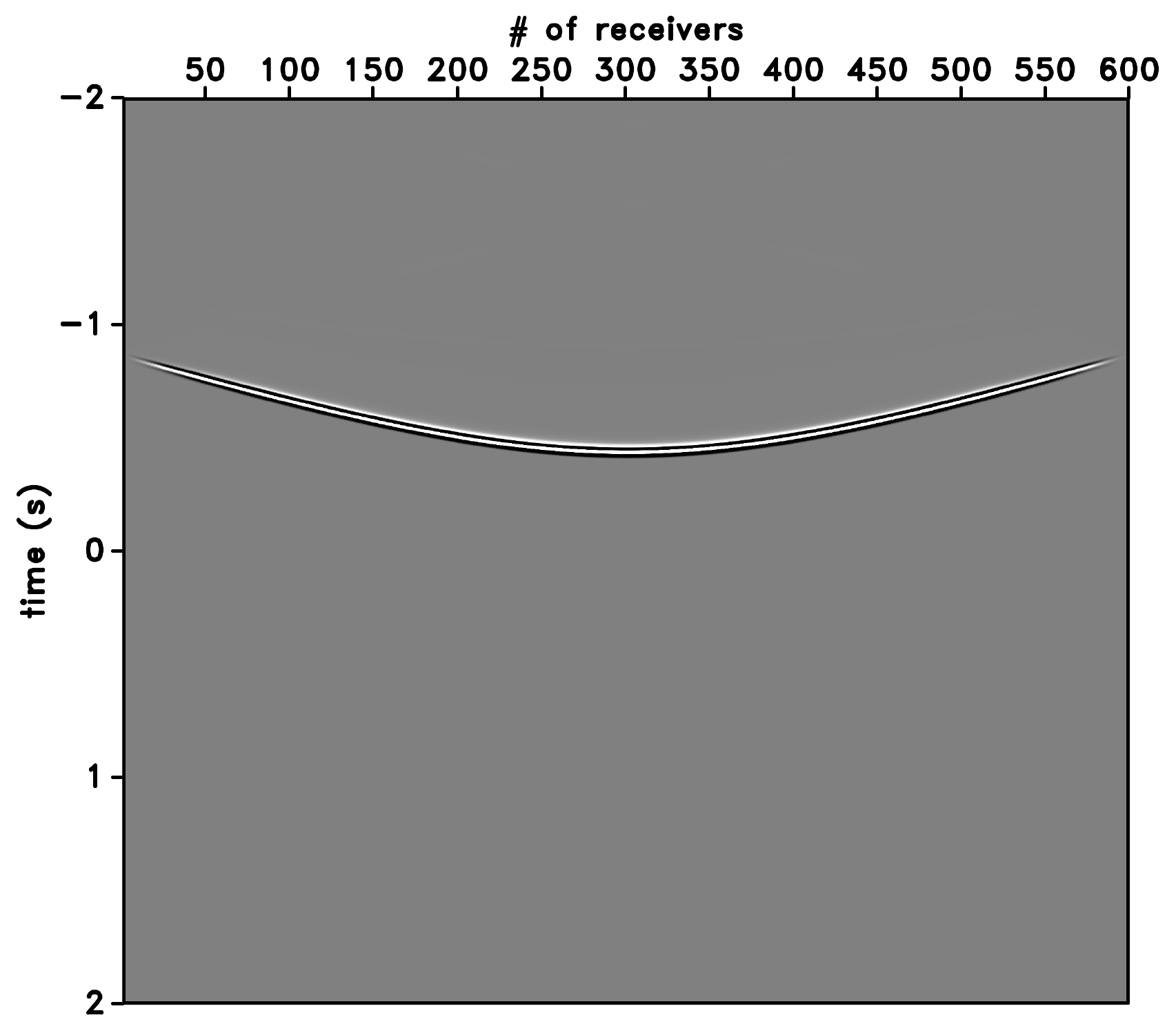}\label{fig:it2}}
  \subfigure[]{\includegraphics[width=0.4\textwidth]{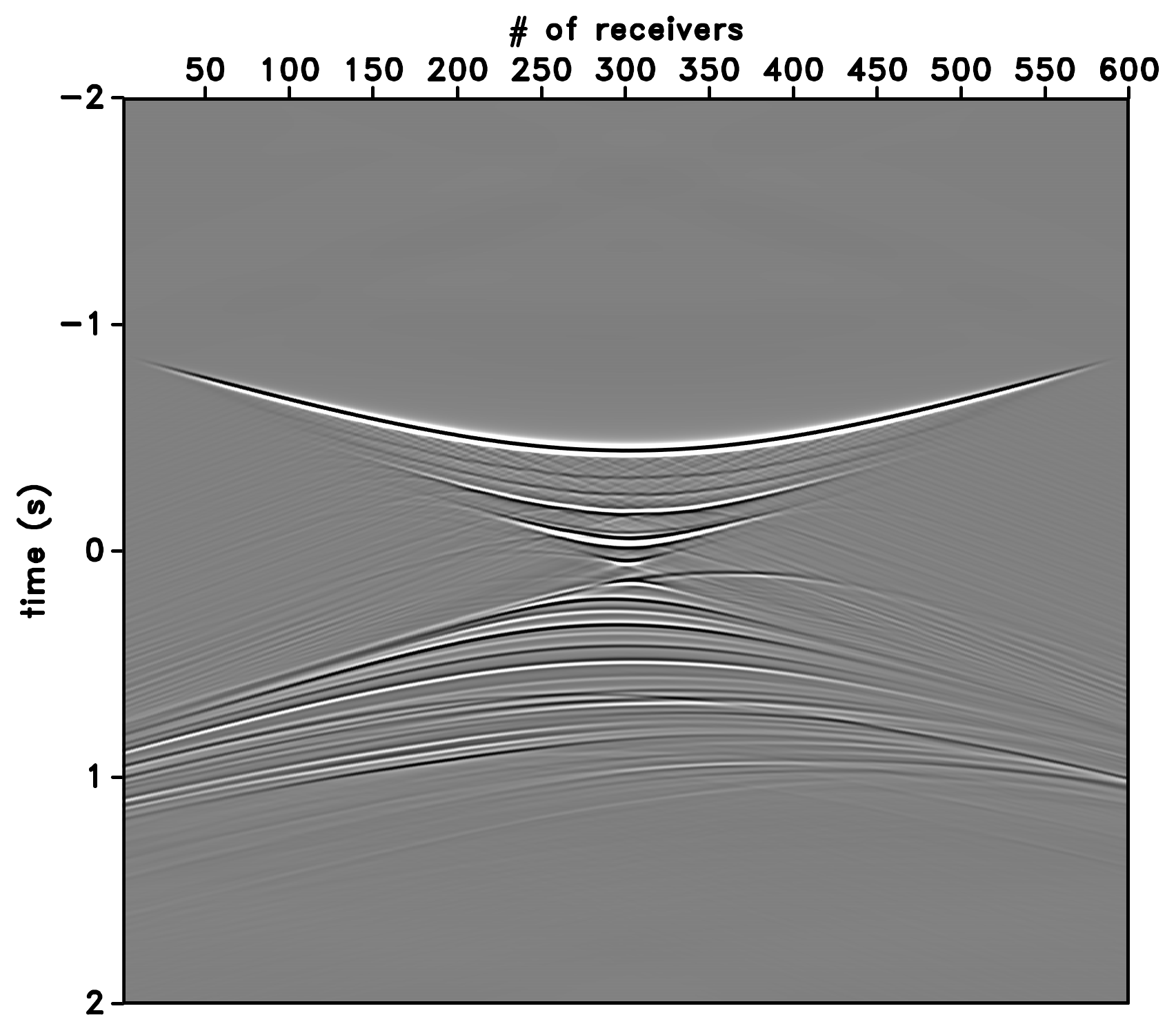}\label{fig:it3}}
  \subfigure[]{\includegraphics[width=0.4\textwidth]{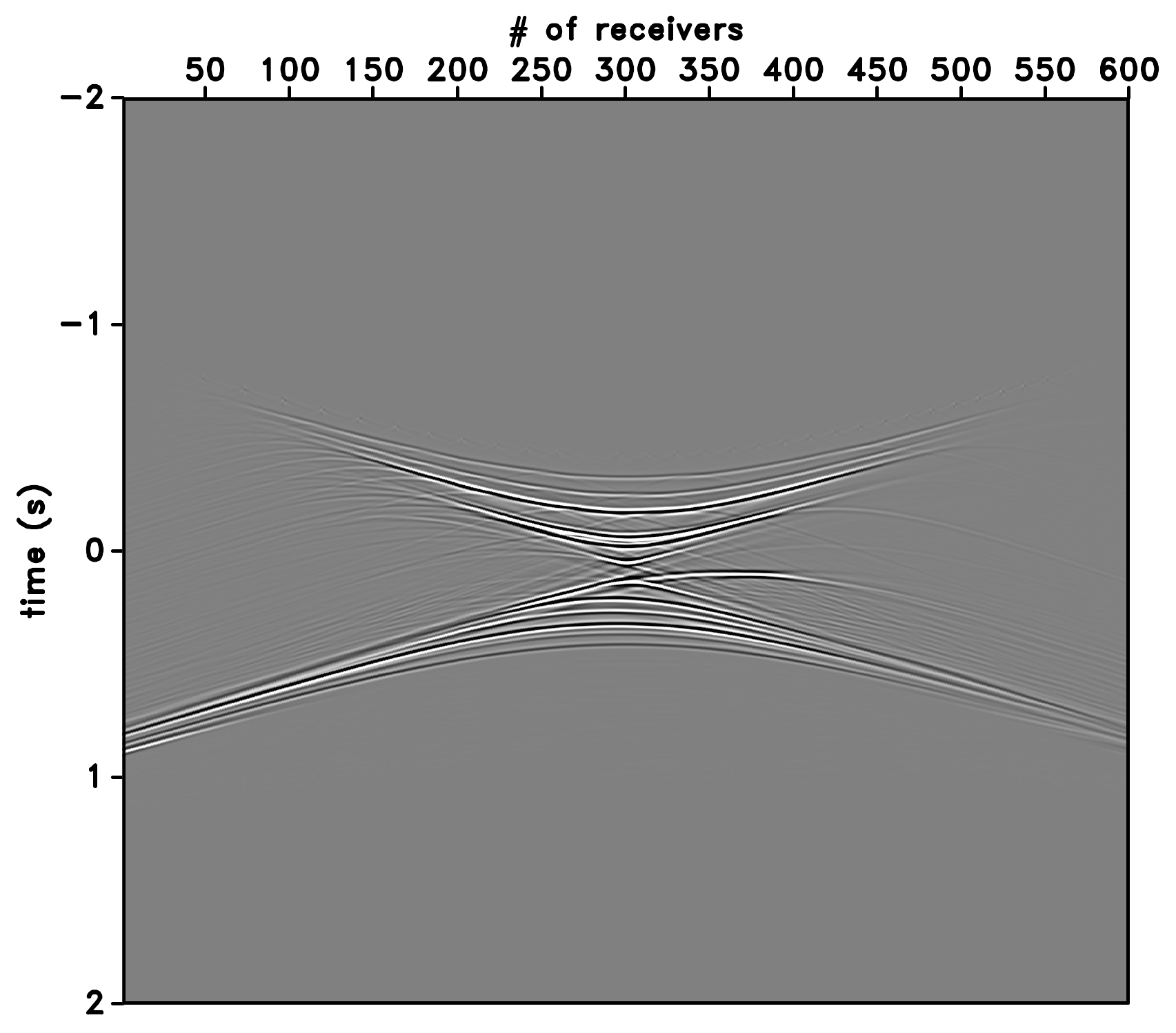}\label{fig:it4}}
  \subfigure[]{\includegraphics[width=0.4\textwidth]{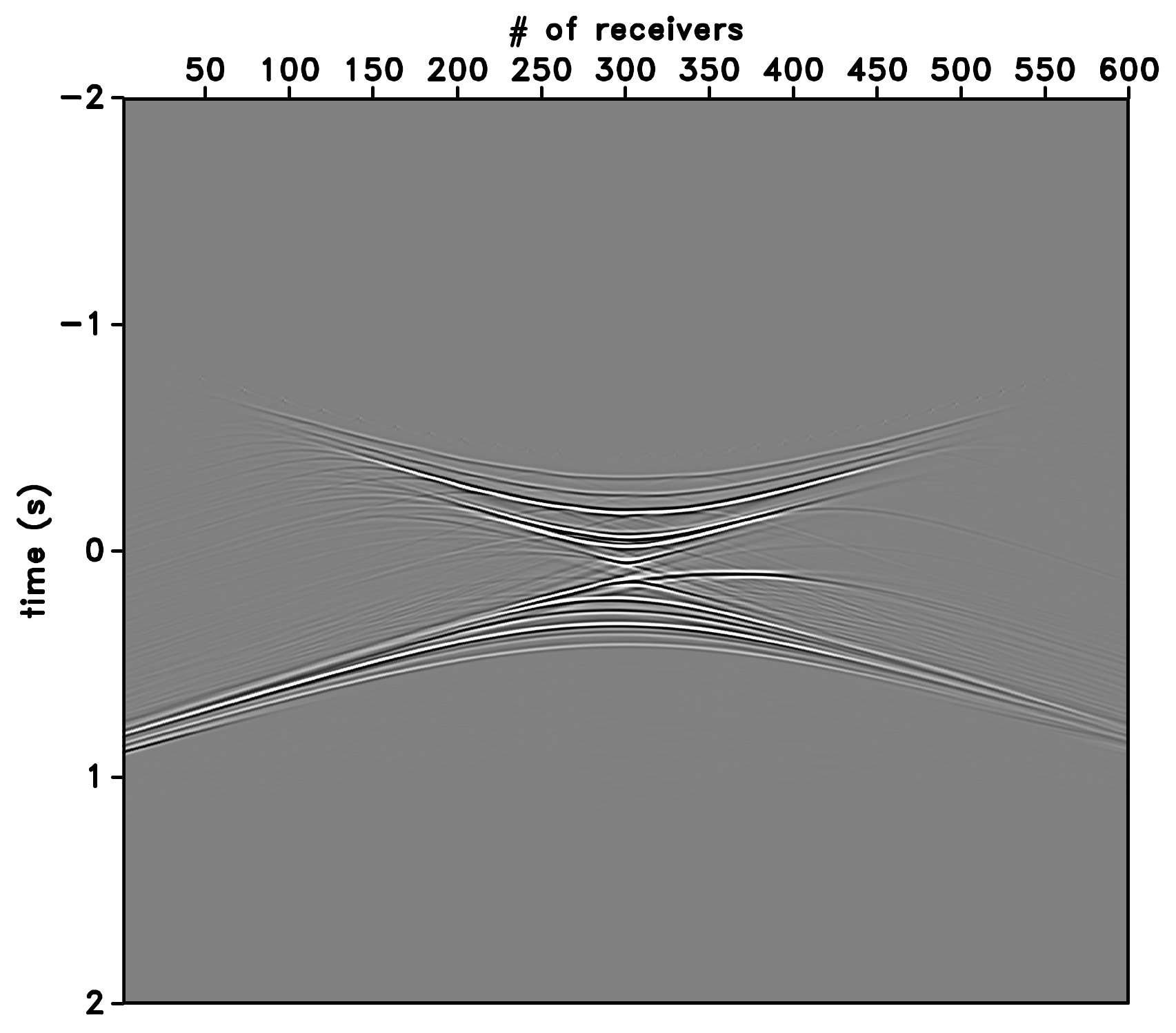}\label{fig:it5}}
  \subfigure[]{\includegraphics[width=0.4\textwidth]{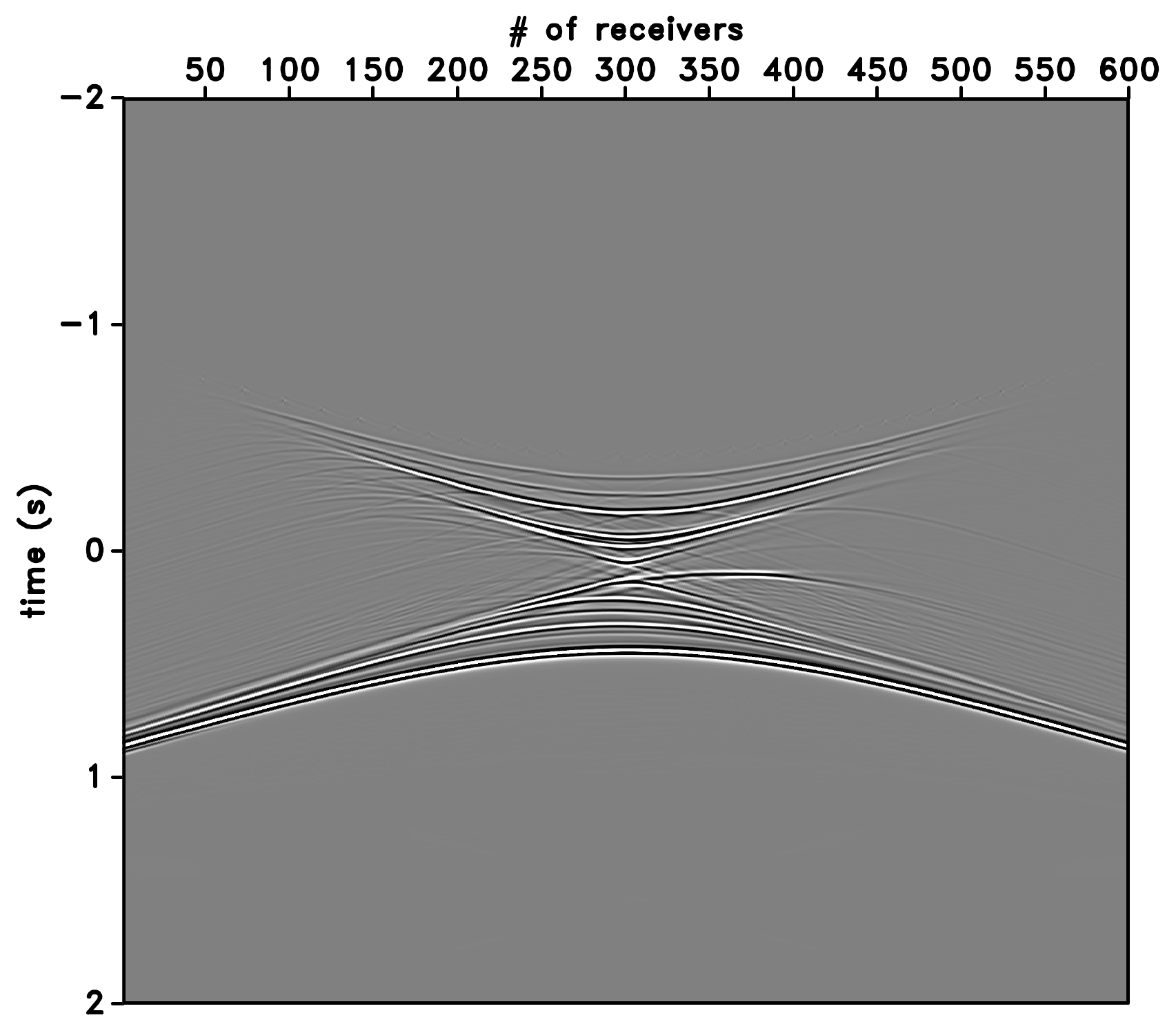}\label{fig:it6}}
  \caption{(a) Modeled direct wave. (b) Time-reversed direct wave. (c) Recorded wavefield, $U_{total}(\textbf{x},t)$, for the first iteration after sending in the time-reversed direct wave. (d) Wavefield after windowing applied. (e) Wavefield after multiplying with (-1). (f) Wavefield after adding the modeled direct wave. This is also the input for the second iteration.}
  \label{fig:iterative}
\end{figure}

\begin{figure}
  \centering
  \subfigure[]{\includegraphics[width=0.4\textwidth]{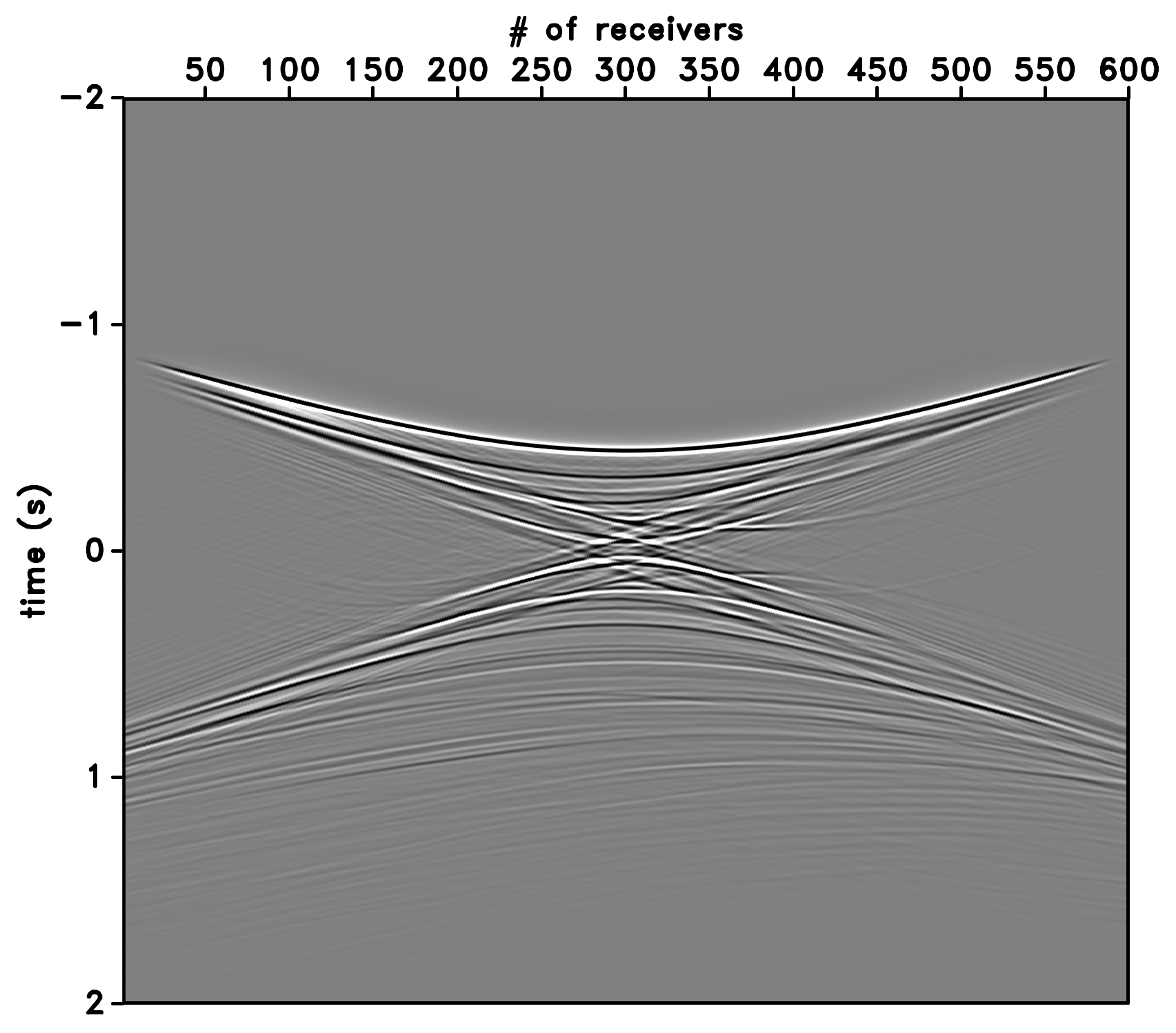}\label{fig:it7}}
  \subfigure[]{\includegraphics[width=0.4\textwidth]{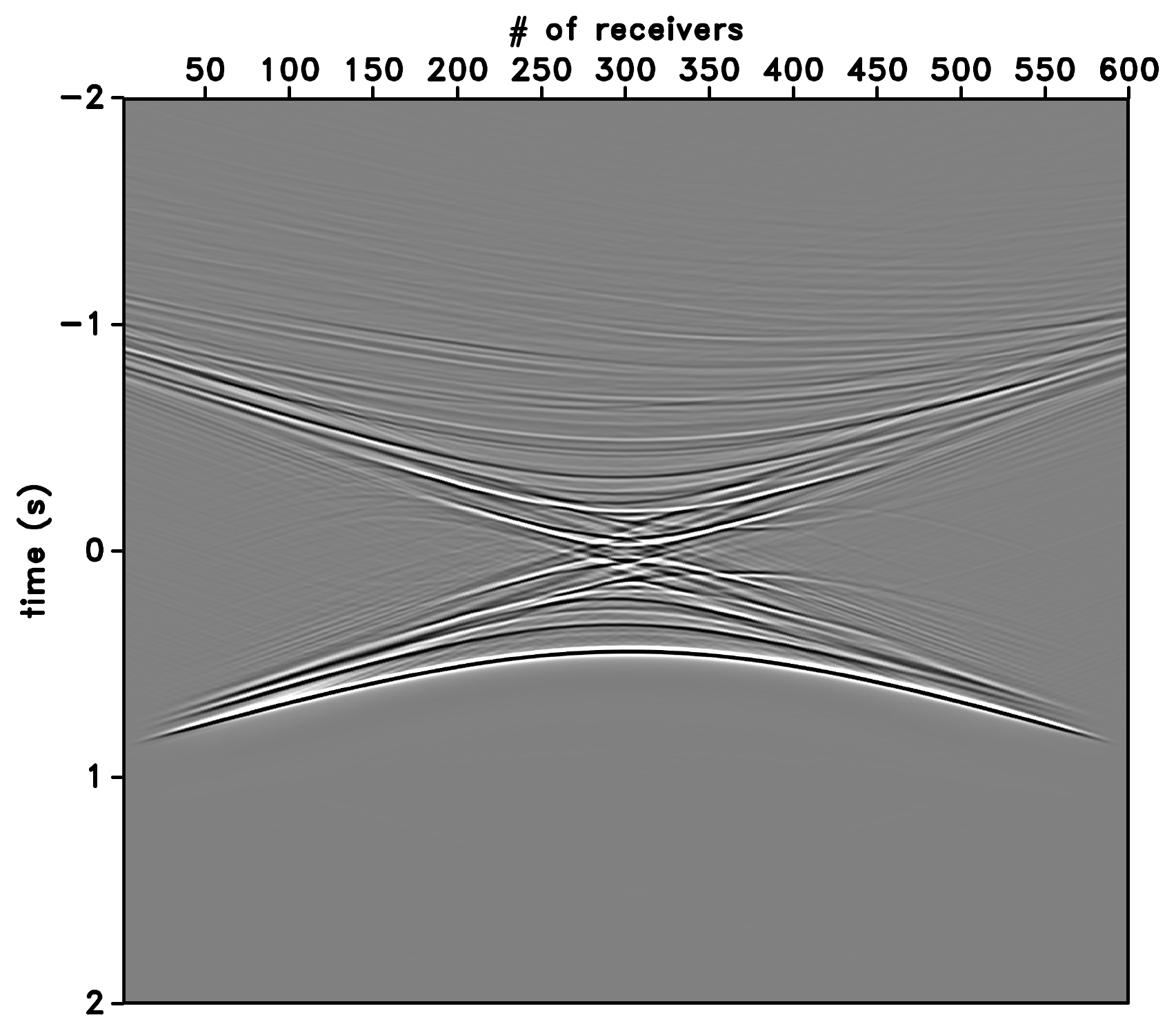}\label{fig:it8}}
  \subfigure[]{\includegraphics[width=0.4\textwidth]{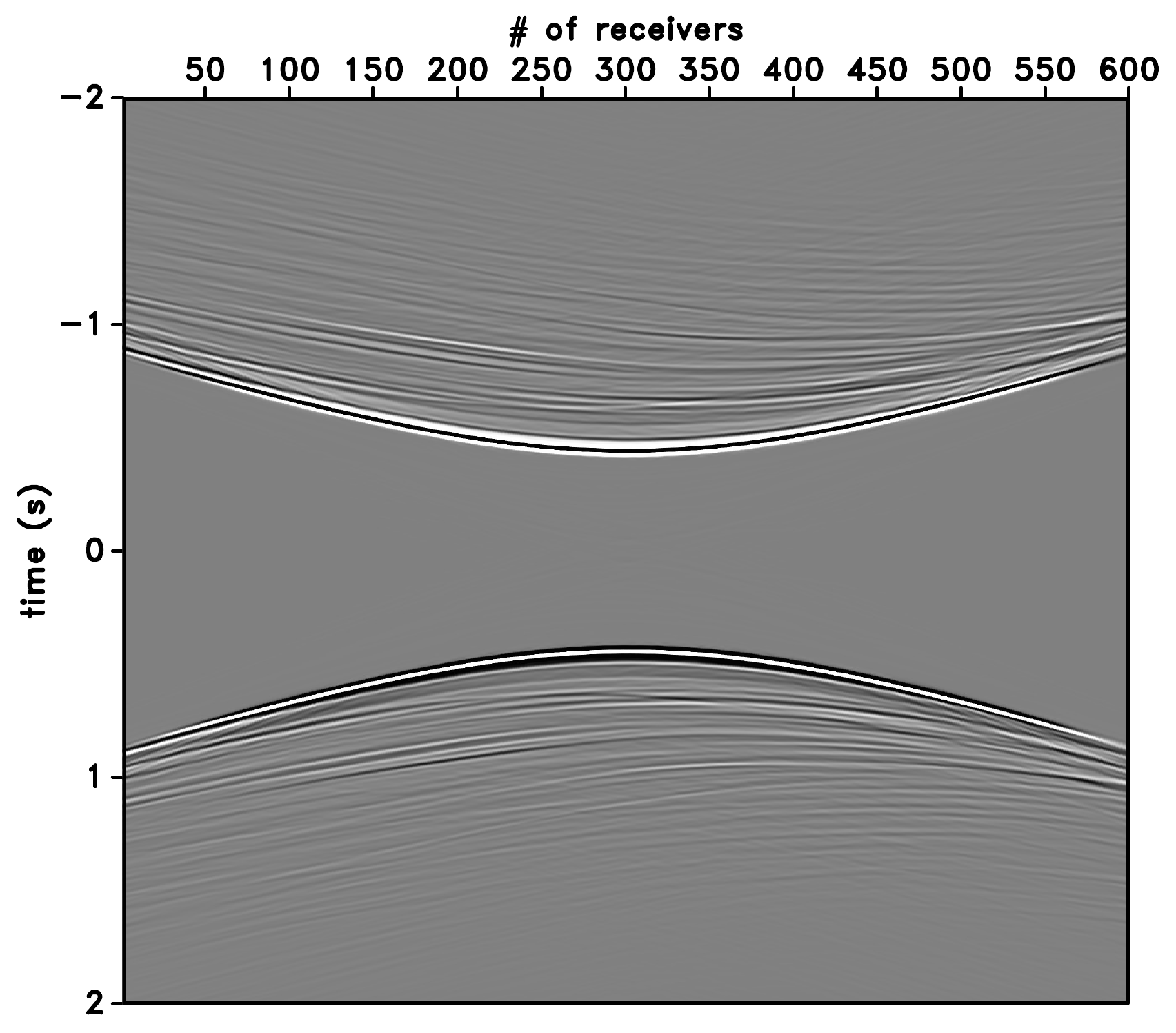}\label{fig:it9}}
  \caption{(a) $U_{total}(\textbf{x},t)$ for the fourth iteration. (b) $U_{total}(\textbf{x},-t)$ for the fourth iteration. (c) $G_{h}(\textbf{x},\textbf{x}_{s},t) = p(\textbf{x},t) - p(\textbf{x},-t)$ for the fourth iteration.}
  \label{fig:homgreen}
\end{figure}

Figure \ref{fig:velsmooth1} shows the smoothed version of the velocity model which is obtained by smoothing the slowness. This model is so strongly smoothed that it presents mostly one-dimensional information about the velocity. As we follow the iterative scheme described in Section 2, the first step is to model the direct wave using the smooth background velocity model (shown in Figure \ref{fig:velsmooth1}). Figure \ref{fig:it1} shows the modeled direct wave. After modeling the direct wave, the next step is to time-reverse the direct wave which is shown in Figure \ref{fig:it2}. We then inject the time-reversed direct wave on the boundary ({one can also convolve the time-reversed direct wave with the reflection response}). Figure \ref{fig:it3} shows the total wavefield, $U_{total}(\textbf{x},t)$. This is the recorded data at the receiver array after sending in the time-reversed direct wave (Figure \ref{fig:it2}) from the receiver array located at the surface of the medium. We use Figures \ref{fig:it1} and \ref{fig:it2} to define the window function ($\Theta$ in equation~(\ref{eq:highit})). After using this window function, Figure \ref{fig:it4} shows the muted data. Following the iterative algorithm, we next negate the muted data (($-$) sign in equation~(\ref{eq:highit})) and the resulting wavefield is shown in Figure \ref{fig:it5}. As the last step, we add the direct wave (Figure \ref{fig:it1}) to this wavefield ($U^{in}_{0}$ in equation~(\ref{eq:highit})) and Figure \ref{fig:it6} shows the combined wavefield ($U_{k}^{in}$ in equation~(\ref{eq:highit})). The time-reverse version of the wavefield shown in Figure \ref{fig:it6} is, therefore, the input for the second iteration and is ready to be sent back into the medium using the receiver array at the surface. 

Figure \ref{fig:it7} shows $U_{total}(\textbf{x},t)$ for the fourth iteration of the iterative scheme. Note that Figure \ref{fig:it7} is nearly symmetric in time for the times $-t_{d} < t < t_{d}$, defined using the arrival time of the direct wave (approximately between -1s and 1s). Figure \ref{fig:it8} shows $U_{total}(\textbf{x},-t)$ for the fourth iteration which is the time-reversed recorded data at the receivers after the fourth iteration. Figure \ref{fig:it9} shows the homogeneous Green's function, $G_{h}(\textbf{x},\textbf{x}_{s},t)$ (see equation (\ref{eq:orishom})), for the fourth iteration after muting the waves between the direct arrivals. 

\begin{figure}
  \centering
  \subfigure[]{\includegraphics[width=0.4\textwidth]{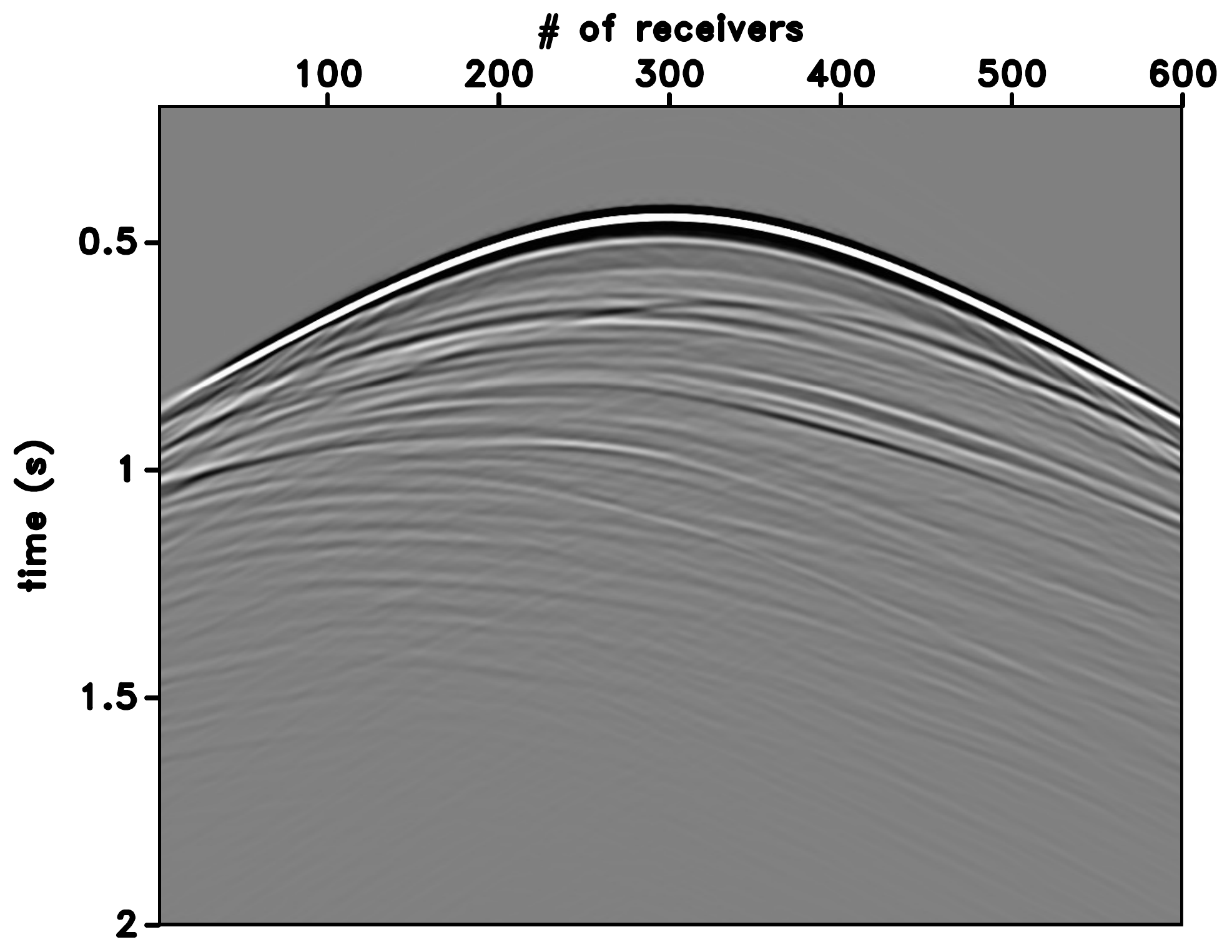}\label{fig:retr}} \\
  \subfigure[]{\includegraphics[width=0.4\textwidth]{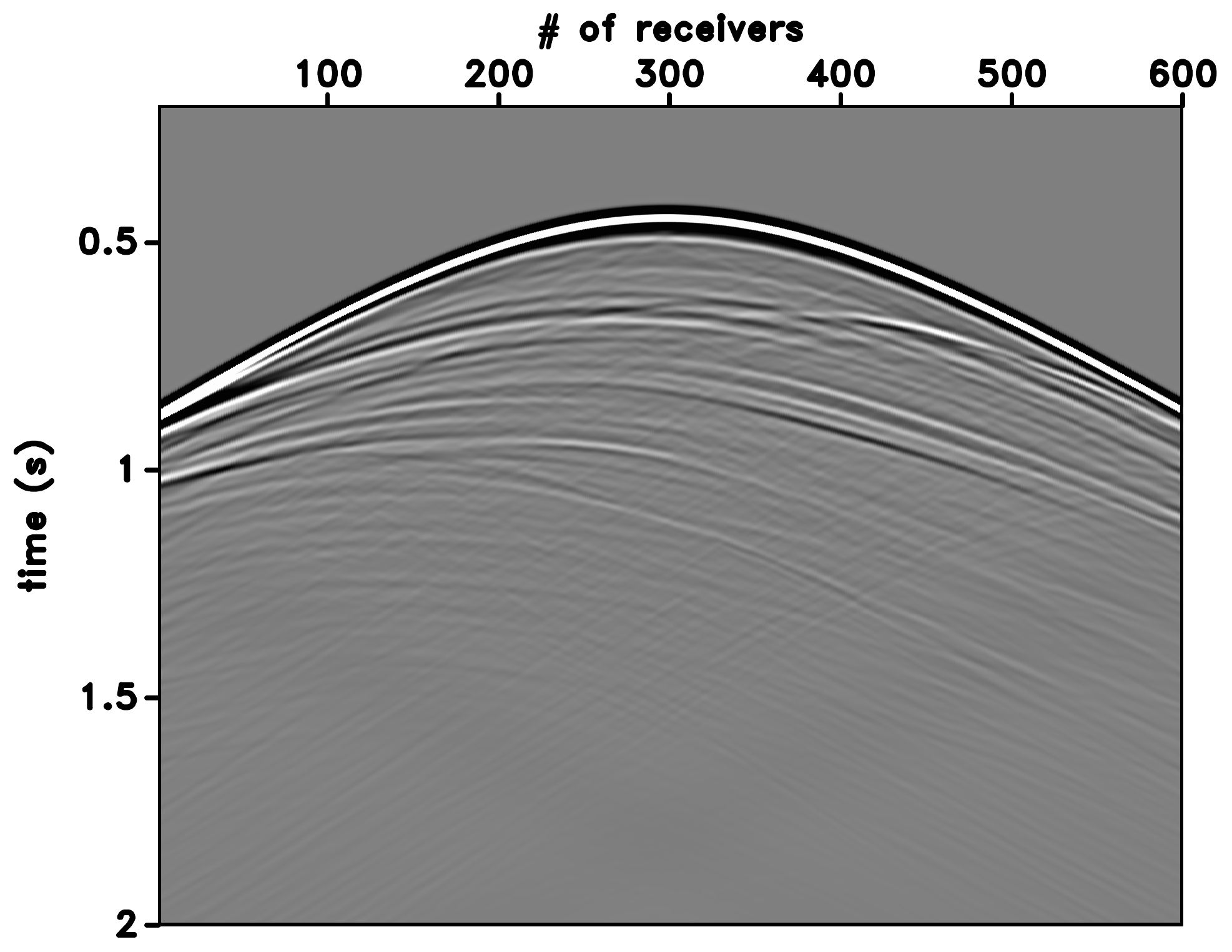}\label{fig:gf}} \\
  \subfigure[]{\includegraphics[width=0.4\textwidth]{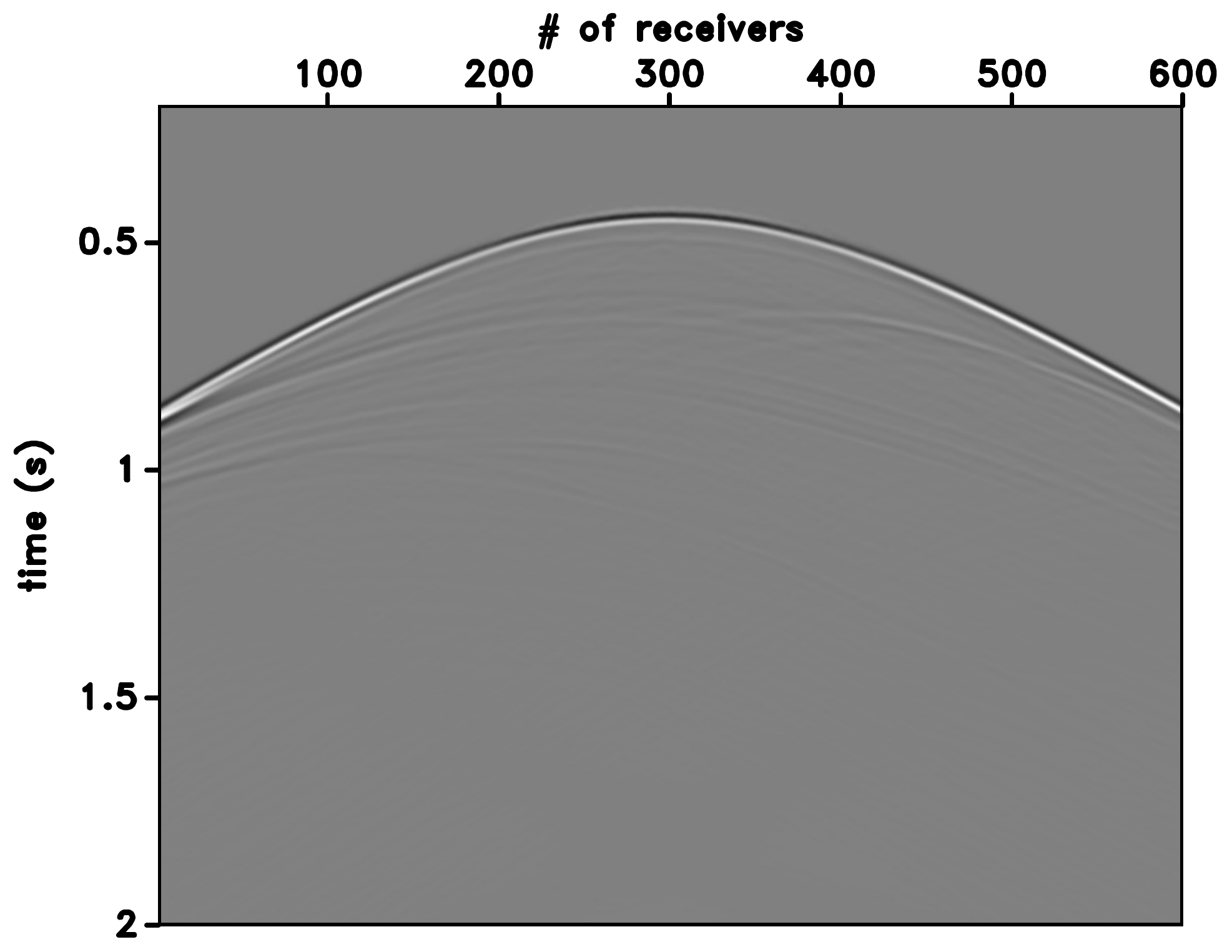}\label{fig:dif-smooth}}
  \caption{(a) Retrieved Green's function using the Marchenko focusing (times when $t>0$ of $G_{h}$ in Figure \ref{fig:it9}). (b) Numerically modeled Green's function. (c) Difference between the numerically modeled Green's function in (a) and the retrieved Green's function in (b).}
  \label{fig:gfandretr}
\end{figure}

Figure \ref{fig:retr} shows $G_{h}(\textbf{x},\textbf{x}_{s},t)$ for the fourth iteration for positive times only which is the retrieved Green's function and Figure \ref{fig:gf} shows the numerically modeled Green's function for the virtual source location. The difference between the numerically modeled and retrieved Green's functions is shown in Figure \ref{fig:dif-smooth}. Figure \ref{fig:gfandretr} shows that the numerically modeled Green's function closely matches the retrieved Green's function for $t \ge t_{d}$. However, we also see in Figure \ref{fig:dif-smooth} that there are overall mismatches in amplitudes and right and left edges of the wavefield that are due to the limited aperture used during the injection of the wavefield back into the medium from the receiver array.

We measure the accuracy of the Green's function retrieval using the Marchenko equation results by calculating trace-by-trace CCs between the retrieved Green's function and the numerically modeled Green's function. The CCs between the retrieved Green's function (Figure \ref{fig:retr}) using the smoothed version of the velocity model and the numerically modeled Green's function (Figure \ref{fig:gf}) are shown in Figure \ref{fig:ccsmooth}. The average CC in Figure \ref{fig:ccsmooth} is 0.76. The low CCs around the right and left edges of the CC plot in Figure \ref{fig:ccsmooth} are due to the limited aperture used during the injection of the wavefield. For the receivers where the limited aperture effects are not evident (receivers from 100 to 500), the average CC is 0.91. This shows a high accuracy Green's function retrieval by the Marchenko focusing algorithm using the smoothed version of the velocity model (Figure \ref{fig:velsmooth1}).

\section{Importance of the initial background velocity model}

\begin{figure}
  \centering
  \includegraphics[width=0.6\textwidth]{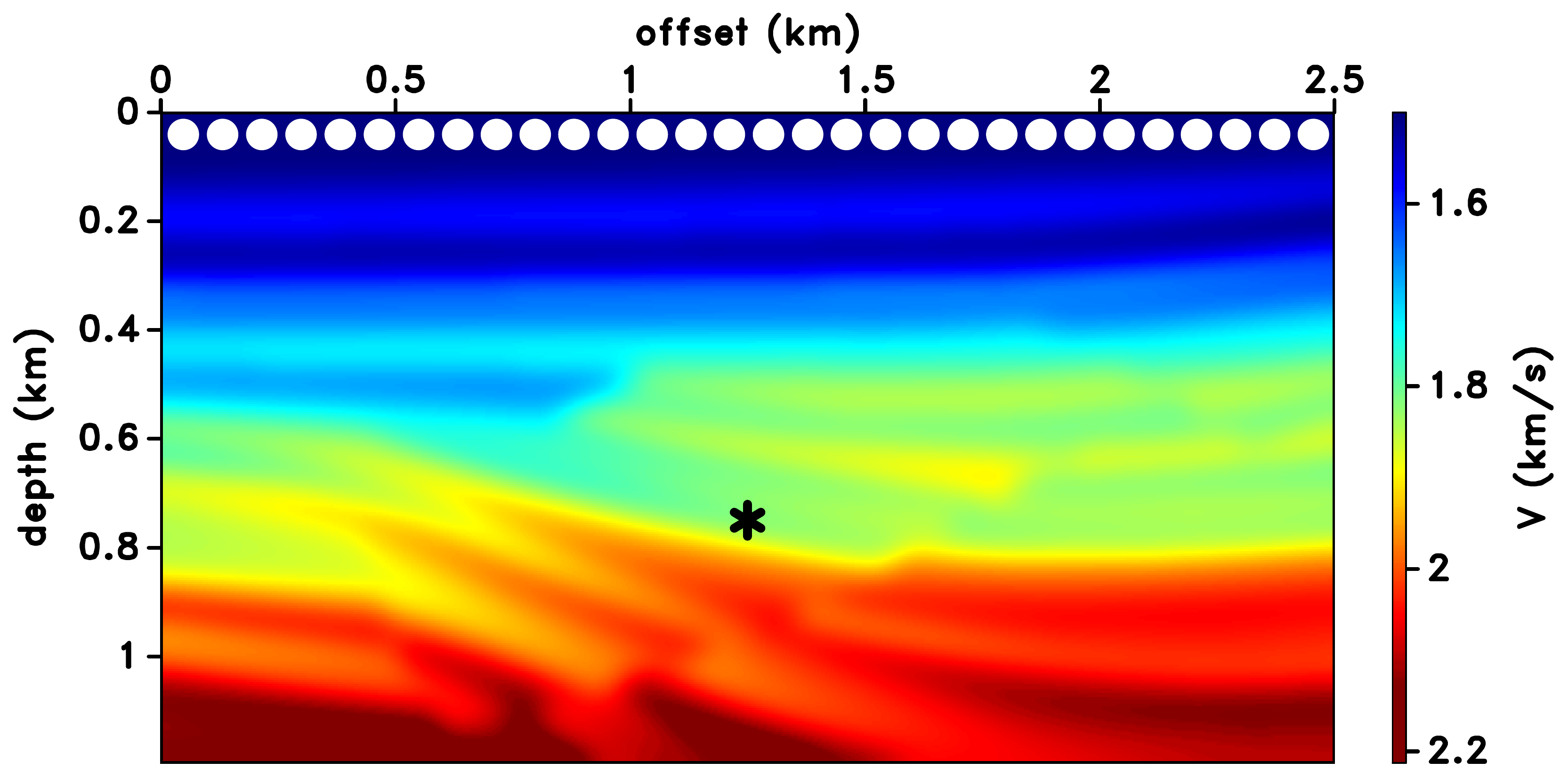}
  \caption{ Less smoothed version of the velocity model used for the iterative algorithm. The black asterisk shows the virtual source location and the white dots at the top indicate every 30th source/receiver location.}
  \label{fig:velsmooth2}
\end{figure}

\begin{figure}
  \centering
  \subfigure[]{\includegraphics[width=0.4\textwidth]{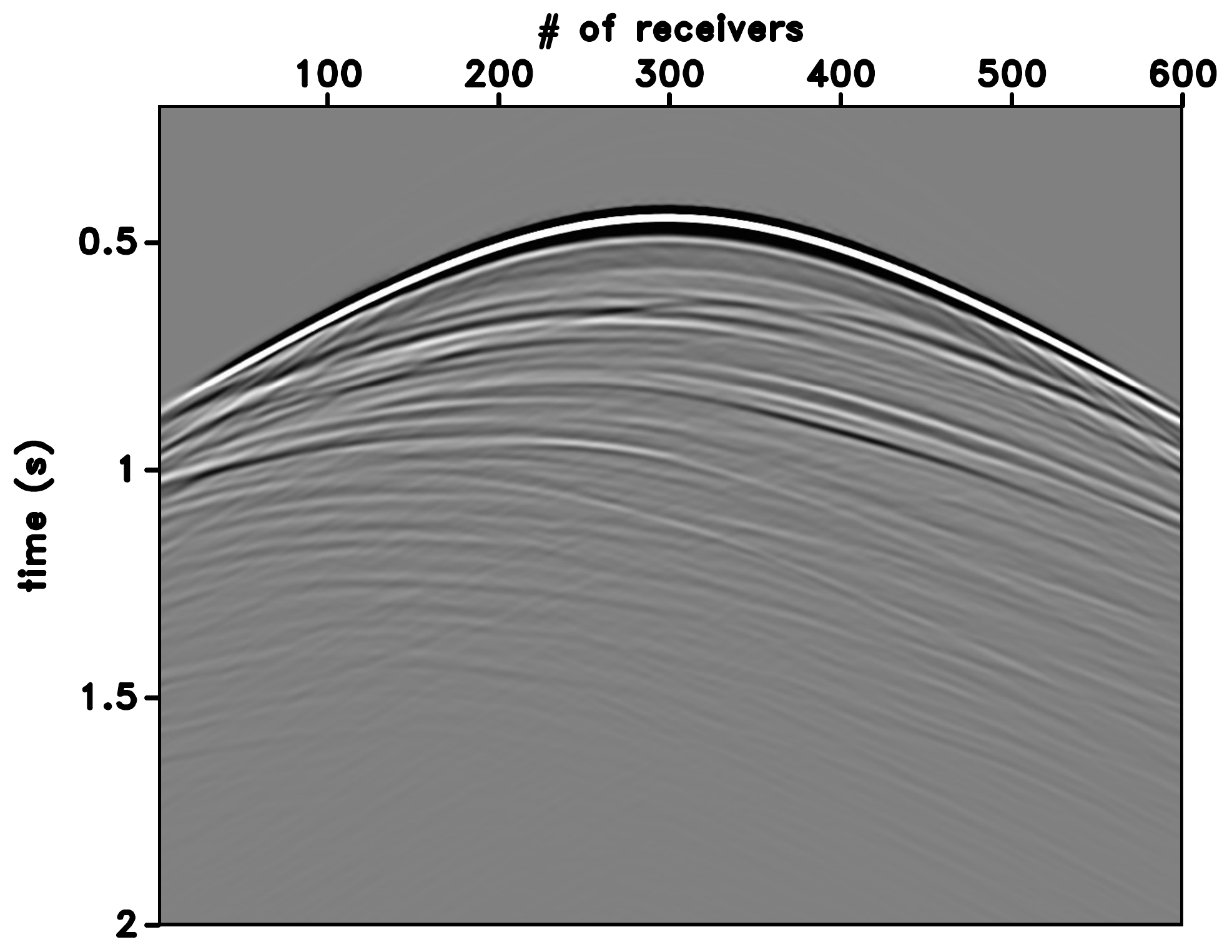}\label{fig:retr-less}} \\
  \subfigure[]{\includegraphics[width=0.4\textwidth]{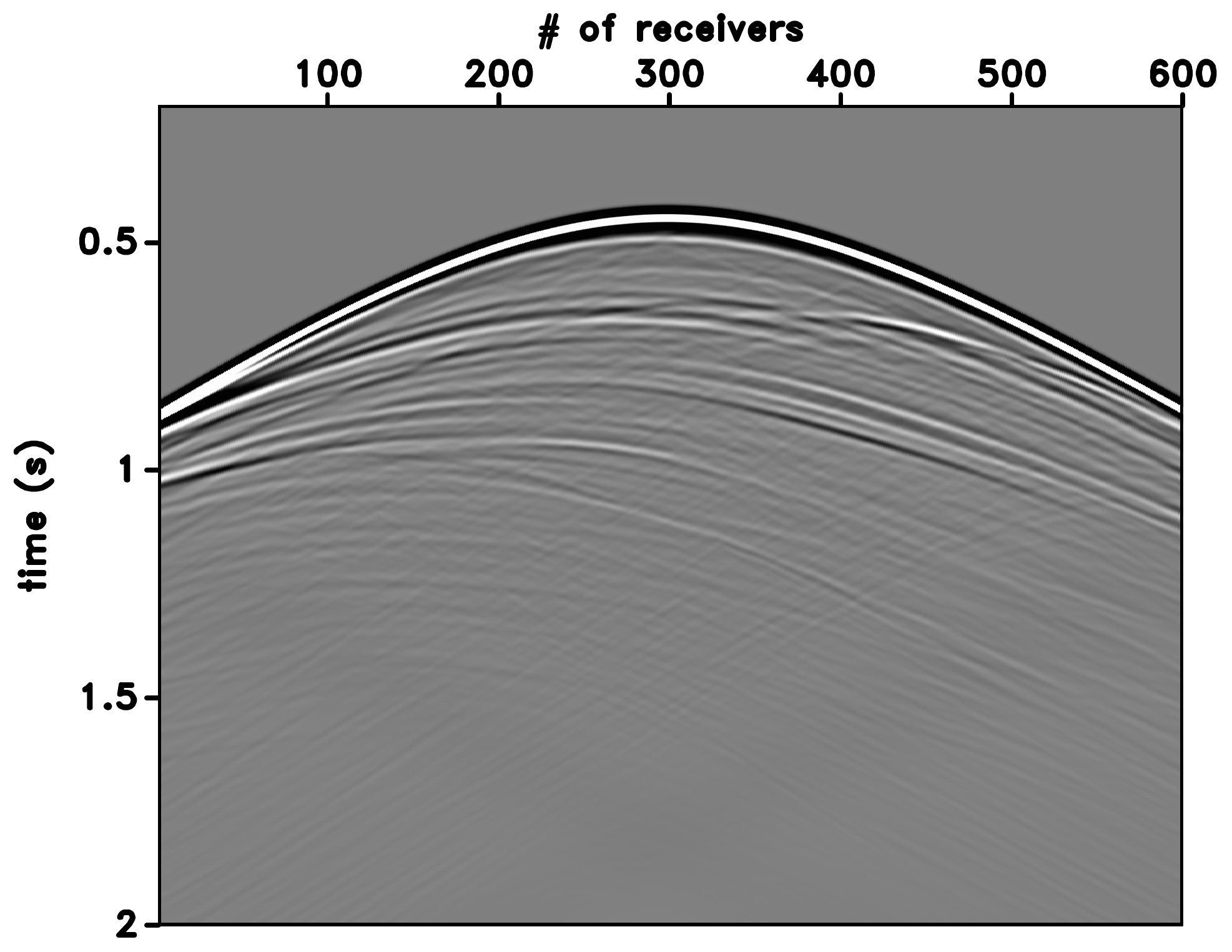}\label{fig:gf-less}} \\
  \subfigure[]{\includegraphics[width=0.4\textwidth]{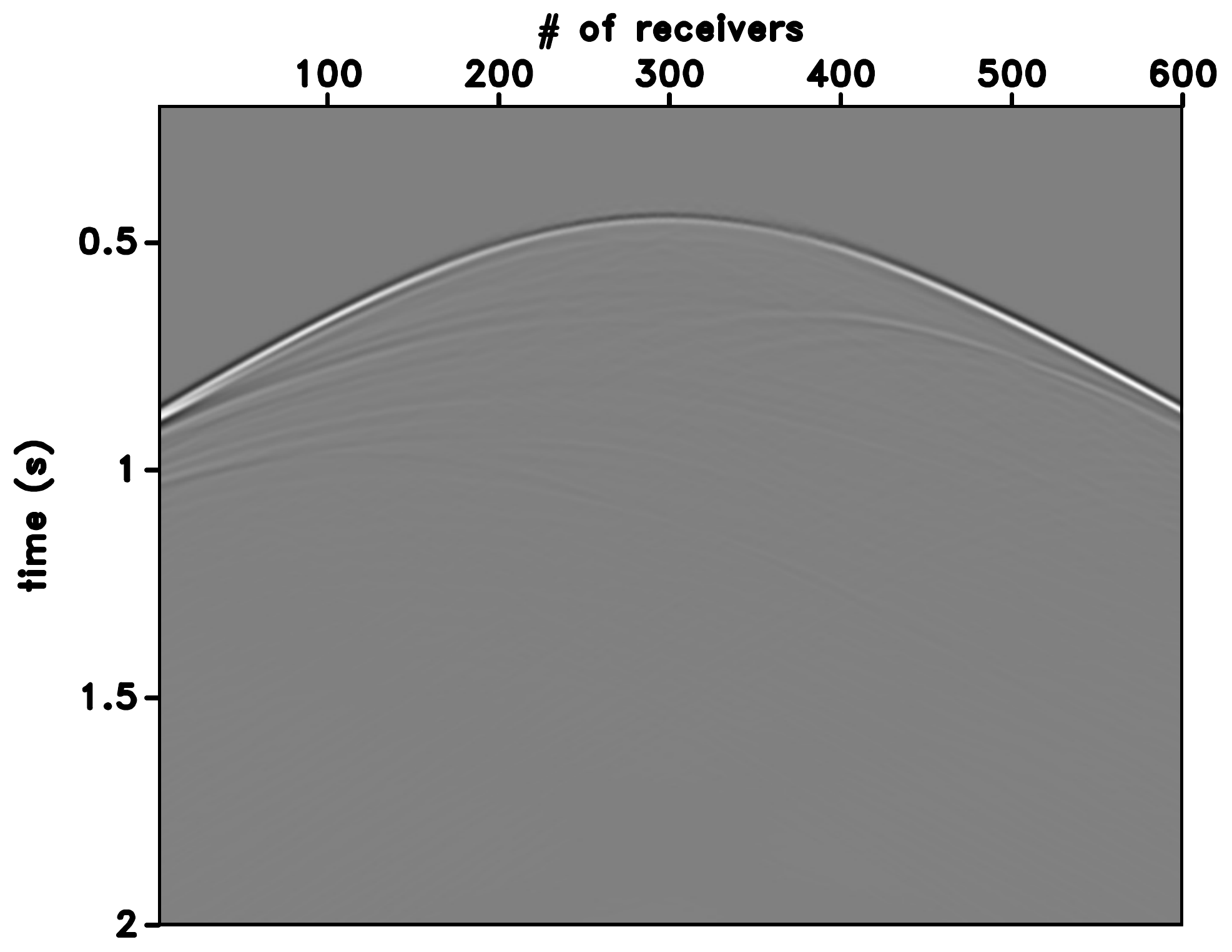}\label{fig:dif-lesssmooth}}
  \caption{(a) Retrieved Green's function using the Marchenko focusing using the less smooth velocity model. (b) Numerically modeled Green's function (which is the same wavefield as Figure \ref{fig:gf}). (c) Difference between the numerically modeled Green's function in (a) and the retrieved Green's function in (b).}
  \label{fig:gfandretr-less}
\end{figure}

In this section, we investigate the effect of the smoothness of the background velocity model on the retrieved Green’s function. The second numerical example consists of the same velocity and density models as with the first example (see Figures \ref{fig:vel1} and \ref{fig:den1}); however, this time we use a less smoothed version of the background velocity model than the one presented in Figure \ref{fig:velsmooth1} to retrieve the Green's function using the Marchenko focusing. Figure \ref{fig:velsmooth2} shows a less smoothed velocity model which has more detailed information about the subsurface structures than the one shown in Figure \ref{fig:velsmooth1}. As with the iterative algorithm, we use the less smoothed velocity model to produce the direct wave and initiate the iterative scheme. By following the steps presented in Figure \ref{fig:iterative}, we retrieve the Green's function. Figure \ref{fig:retr-less} shows $G_{h}$ for the fourth iteration for positive times only (being the retrieved Green's function) and Figure \ref{fig:gf-less} shows the numerically modeled Green's function for the virtual source location. The difference between the numerically modeled Green's function (Figure \ref{fig:gf-less}) and the retrieved Green's function Figure \ref{fig:retr-less} is shown in Figure \ref{fig:dif-lesssmooth}. Similar to Figure \ref{fig:dif-smooth}, Figure \ref{fig:dif-lesssmooth} shows that the retrieved and modeled Green's functions have mismatches in overall amplitude, and the right and left wavefield edges.

\begin{figure}
  \centering
  \includegraphics[width=0.6\textwidth]{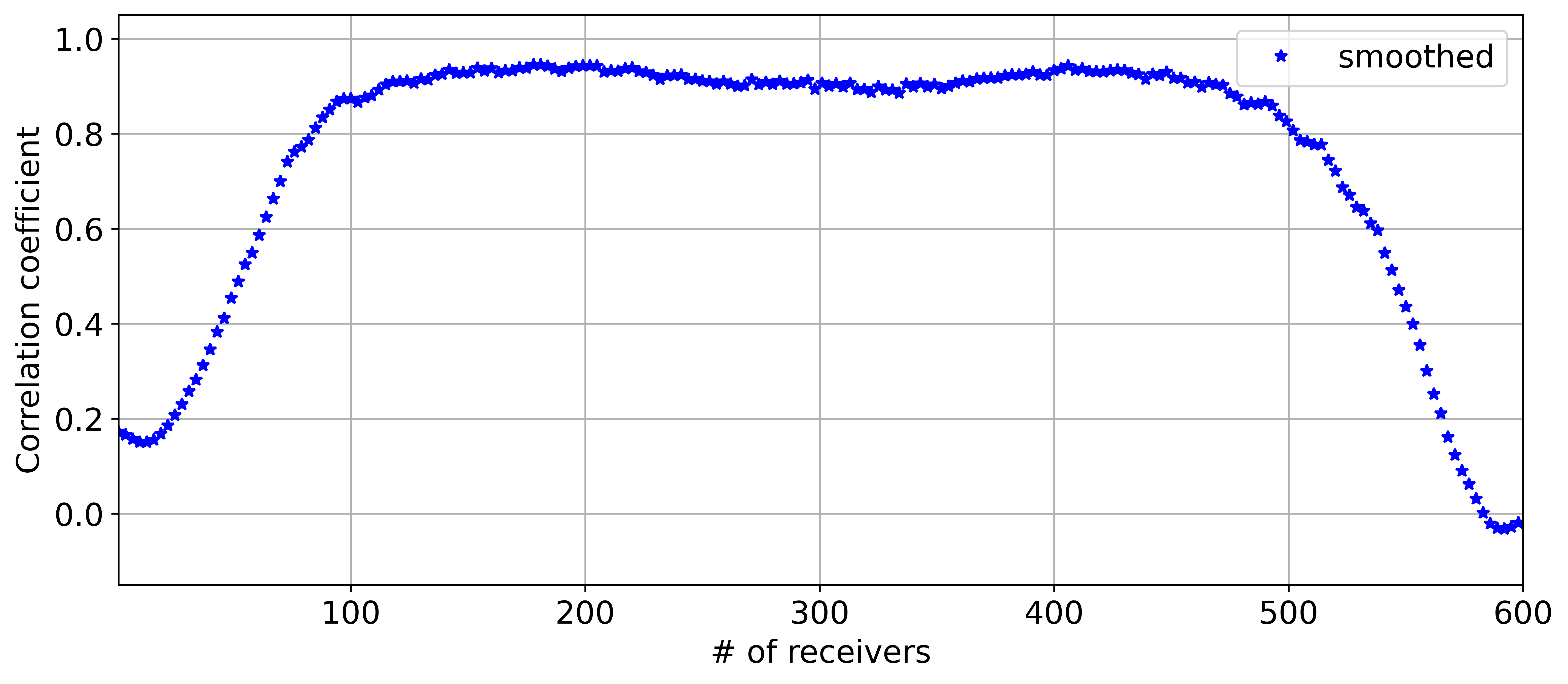}
     \caption{Trace-by-trace calculated correlation coefficient between the retrieved Green's function (Figure \ref{fig:retr}) and the numerically modeled Green's function (Figure \ref{fig:gf}).}
  \label{fig:ccsmooth}
\end{figure}

\begin{figure}
  \centering
   \includegraphics[width=0.6\textwidth]{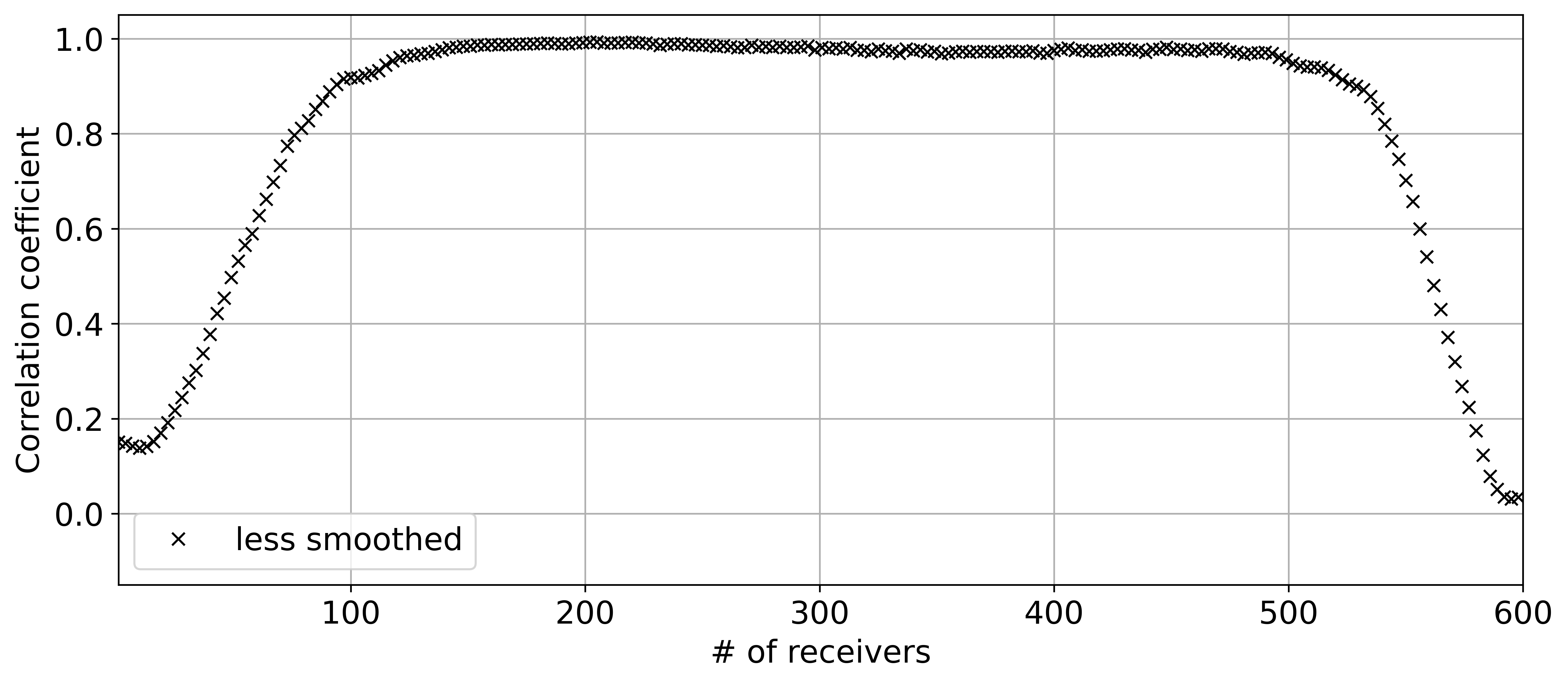}
     \caption{Trace-by-trace calculated correlation coefficient between the retrieved Green's function (Figure \ref{fig:retr-less}) and the numerically modeled Green's function (Figures \ref{fig:gf} and \ref{fig:gf-less}).}
  \label{fig:cclesssmooth}
\end{figure}

To quantify the quality of the retrieved Green's function using the less smoothed velocity model (Figure \ref{fig:velsmooth2}), we calculate the CCs between the retrieved Green's function (Figure \ref{fig:retr-less}) and the numerically modeled Green's function (Figure \ref{fig:gf-less}), which are shown in Figure \ref{fig:cclesssmooth}. The average CC in Figure \ref{fig:cclesssmooth} is 0.83, and the average CC for receivers from 100 to 500 is 0.98. Therefore, by using a less smoothed version of the velocity model (Figure \ref{fig:velsmooth2}) for the iterative scheme, we retrieve a more accurate Green's function by the Marchenko focusing algorithm than the one presented in Figure \ref{fig:ccsmooth}.

\begin{figure}
  \centering
  \includegraphics[width=0.6\textwidth]{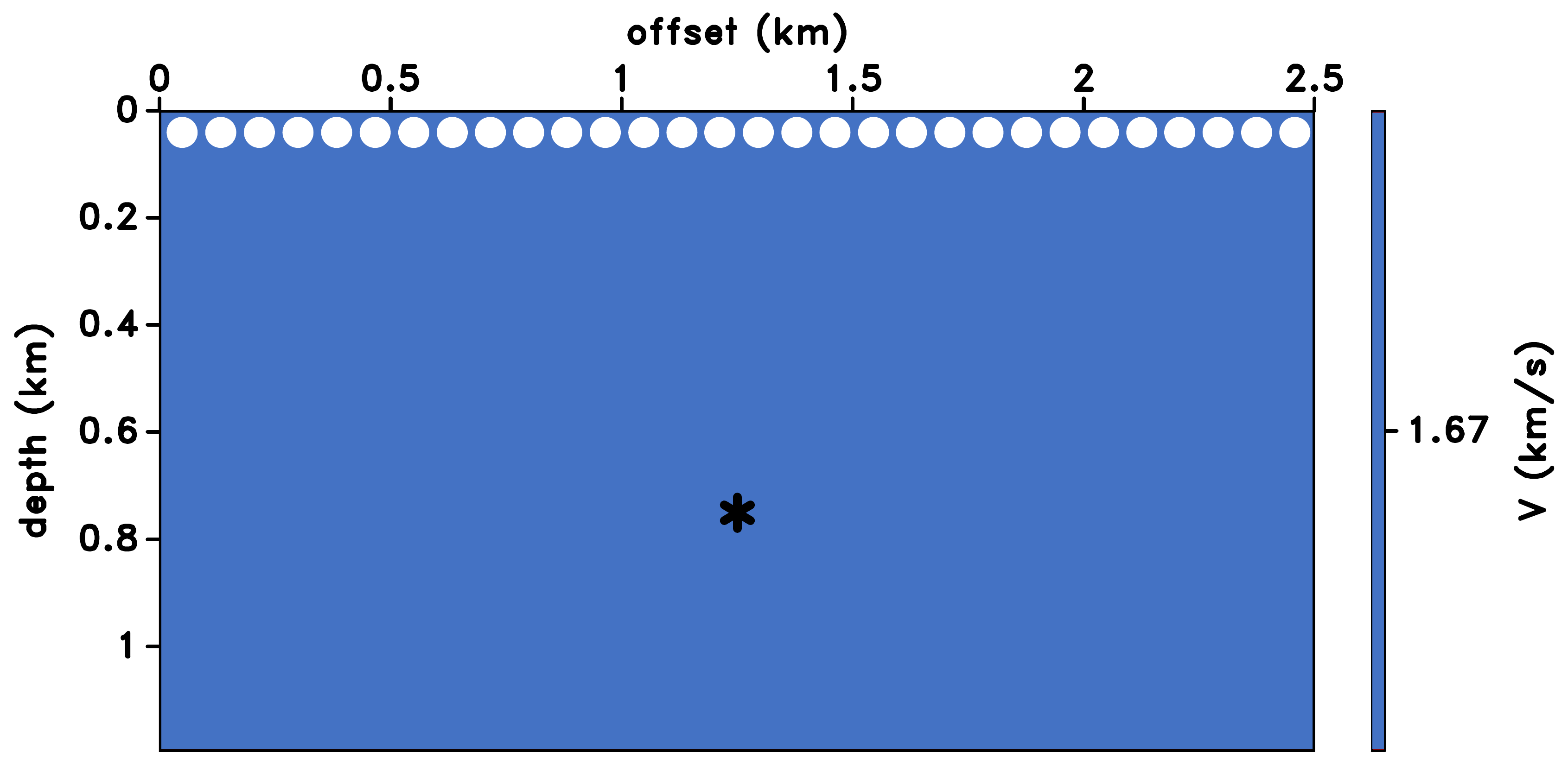}
  \caption{Constant velocity model used for the iterative algorithm. The black asterisk shows the virtual source location and the white dots at the top indicate every 30th source/receiver location.}
  \label{fig:velsmooth3}
\end{figure}

\begin{figure}
  \centering
  \subfigure[]{\includegraphics[width=0.4\textwidth]{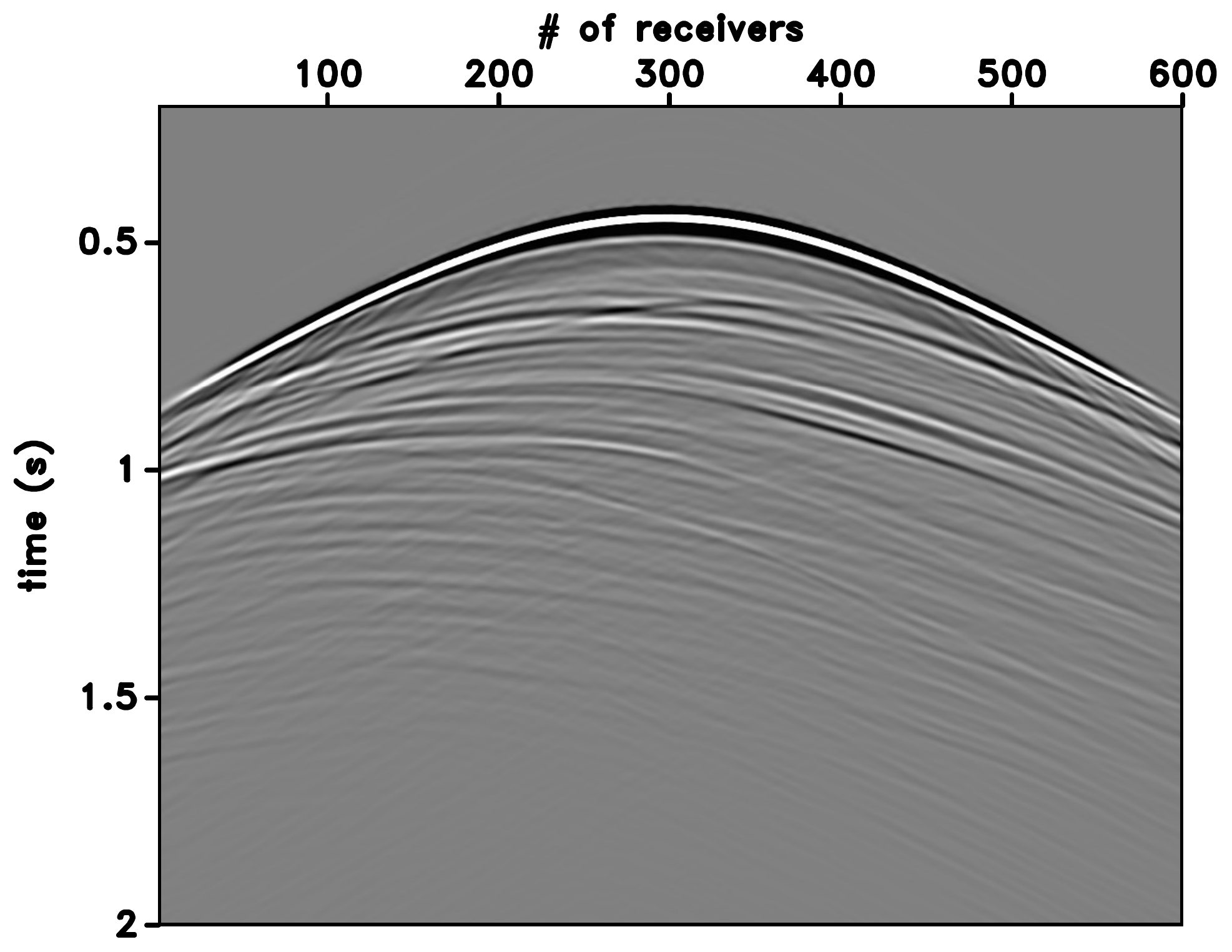}\label{fig:retr-cons}} \\
  \subfigure[]{\includegraphics[width=0.4\textwidth]{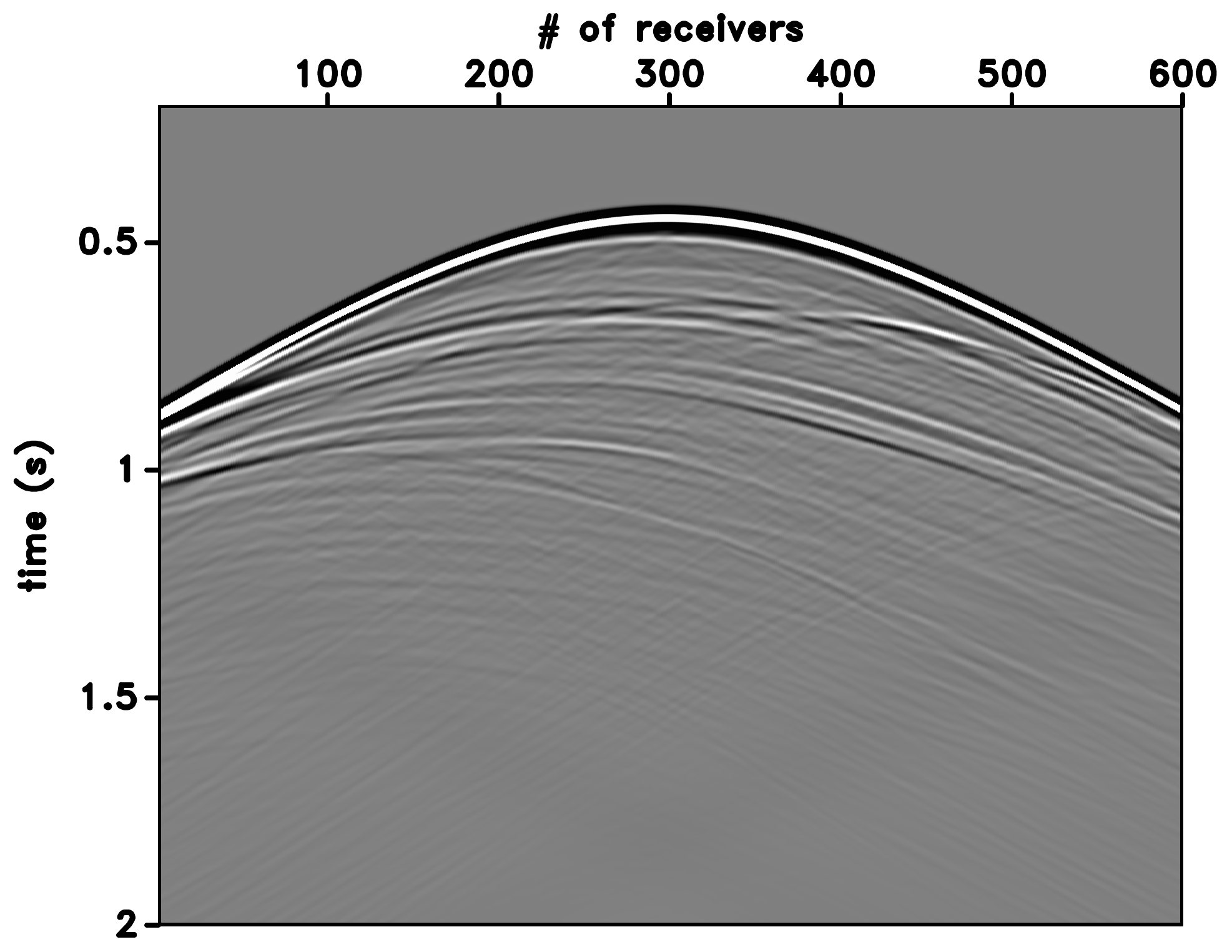}\label{fig:gf-cons}} \\
  \subfigure[]{\includegraphics[width=0.4\textwidth]{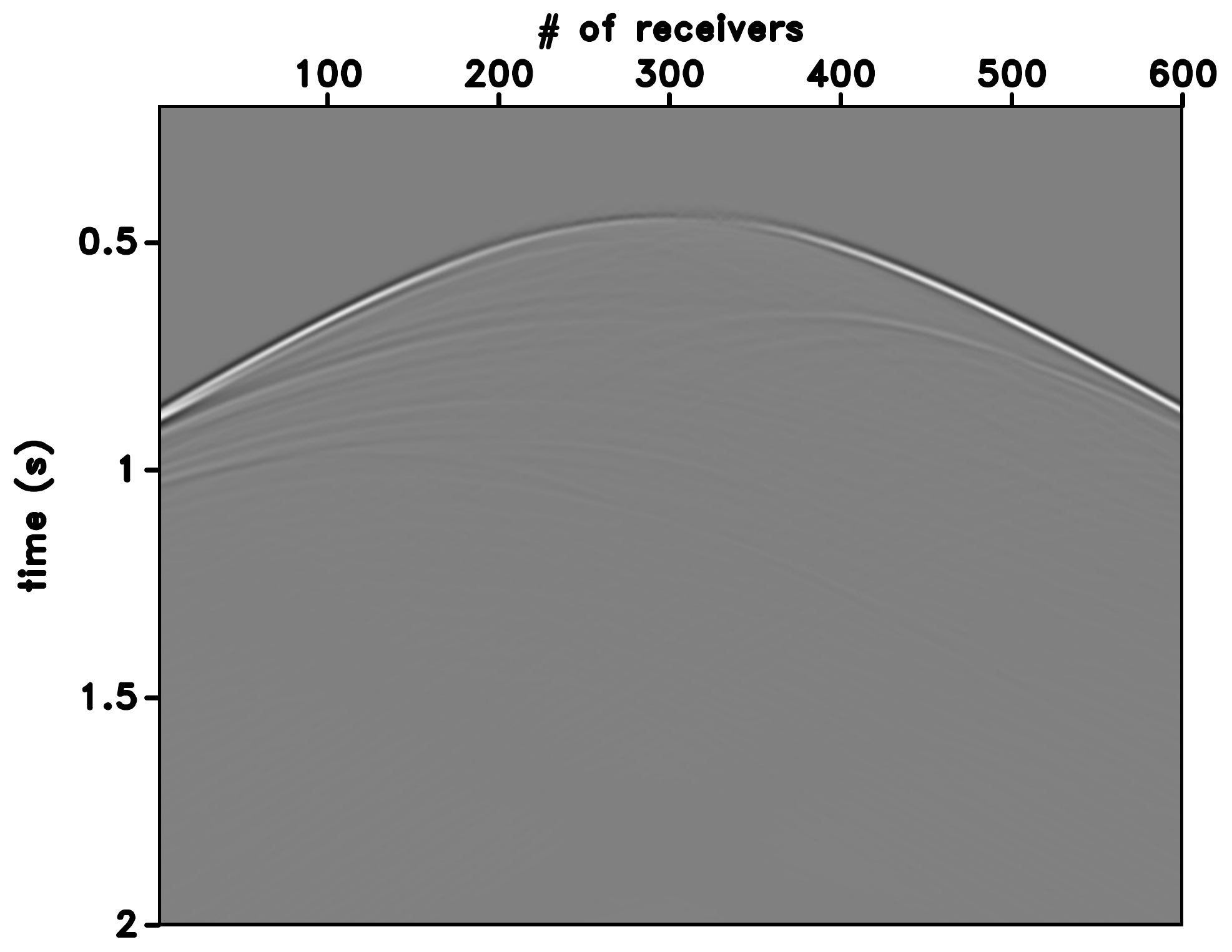}\label{fig:dif-cons}}
  \caption{(a) Retrieved Green's function using the Marchenko focusing using the constant velocity model. (b) Numerically modeled Green's function (which is the same wavefield as Figures \ref{fig:gf} and \ref{fig:gf-less}). (c) Difference between the numerically modeled Green's function in (a) and the retrieved Green's function in (b).}
  \label{fig:gfandretr-cons}
\end{figure}

As the last step of the velocity model sensitivity analysis, we use a constant velocity model. As opposed to the first two velocity models used (see Figures \ref{fig:velsmooth1} and \ref{fig:velsmooth2}), the constant velocity model does not include any geological or geophysical information, including about the possible dipping layers and the velocity variations. The constant value of the velocity is calculated using the average slowness between the surface and the depth of the focal point and is shown in Figure \ref{fig:velsmooth3}, which is used to model the direct wave for the iterative algorithm. After following the iterative scheme, Figure \ref{fig:retr-cons} shows $G_{h}$ for the fourth iteration for positive times only (the retrieved Green's function), and Figure \ref{fig:gf-cons} shows the numerically modeled Green's function for the virtual source location. Figure \ref{fig:dif-cons} shows the difference between the numerically modeled Green's function (Figure \ref{fig:gf-cons}) and the retrieved Green's function (Figure \ref{fig:retr-cons}). The difference between the numerically modeled and the retrieved Green's functions in Figure \ref{fig:dif-cons} using the constant velocity model (Figure \ref{fig:velsmooth3}) is similar to the ones presented in Figures \ref{fig:dif-smooth} and \ref{fig:dif-lesssmooth}. There are also mismatches in overall amplitudes, and the right and left wavefield edges. 

\begin{figure}
  \centering
  \includegraphics[width=0.6\textwidth]{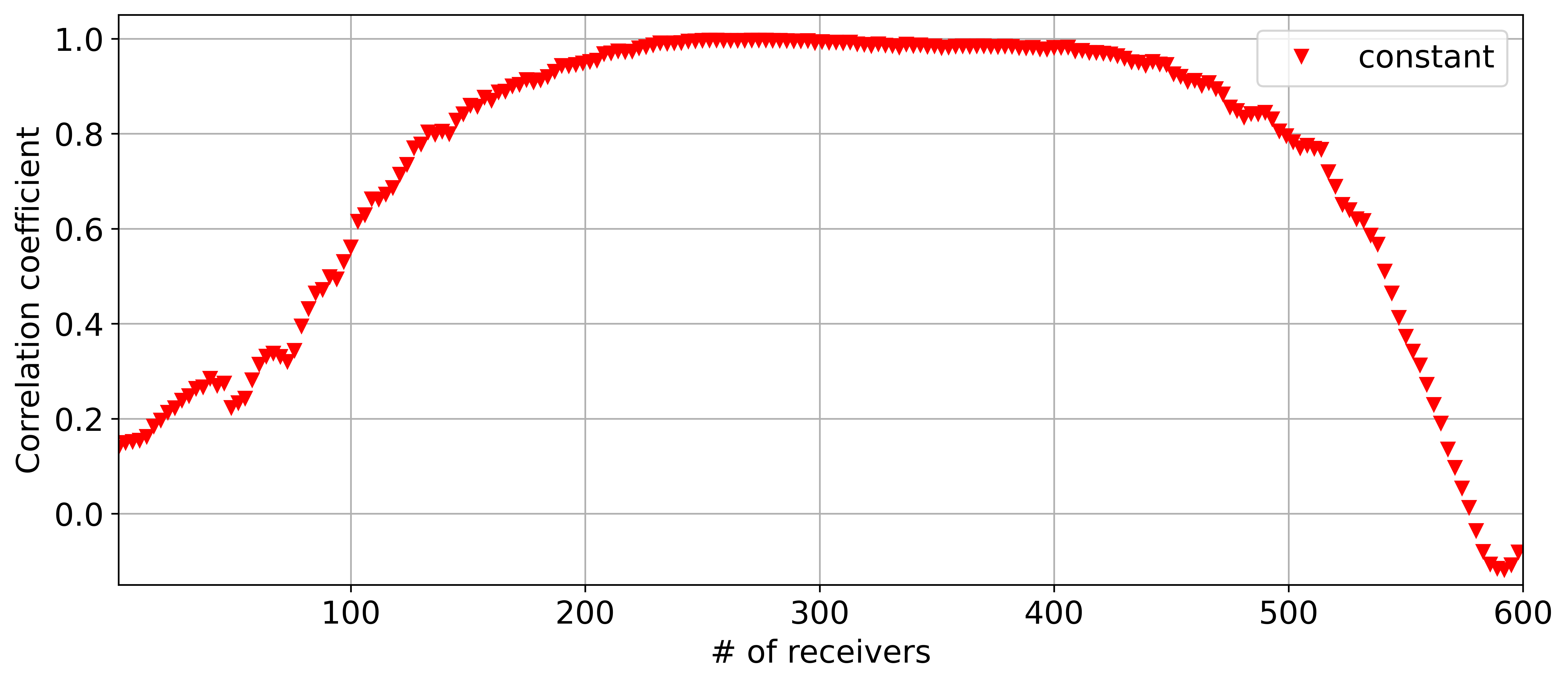}
     \caption{Trace-by-trace calculated correlation coefficient between the retrieved Green's function (Figure \ref{fig:retr-cons}) and the numerically modeled Green's function (Figures \ref{fig:gf}, \ref{fig:gf-less}, and \ref{fig:gf-cons}).}
  \label{fig:cccons}
\end{figure}

Similar to the previous examples, we show the accuracy of the retrieved Green's function using the constant velocity model (Figure \ref{fig:velsmooth3}) by calculating the CCs between the retrieved (Figure \ref{fig:retr-cons}) and the numerically modeled (Figure \ref{fig:gf-cons}) Green's functions, which are shown in Figure \ref{fig:cccons}. The average CC in Figure \ref{fig:cclesssmooth} is 0.73; however, the average CC for receivers from 100 to 500 is 0.93. The CC for receivers from 100 to 500 in Figure \ref{fig:cccons} is higher than the one presented in Figure \ref{fig:ccsmooth} for receivers from 100 to 500; however, the CC for receivers from 0 to 600 in Figure \ref{fig:cccons} is lower than the CC presented in Figure \ref{fig:ccsmooth}. Using a constant velocity model for the iterative scheme, we retrieve just as accurate Green's function as with using the smoothed velocity model for the Marchenko focusing algorithm for receivers close to the virtual source location; however, as the offset (or the horizontal extent of the model) increases, the accuracy of the retrieved Green's function decreases for the constant velocity model. 

\begin{figure}
  \centering
  \includegraphics[width=0.6\textwidth]{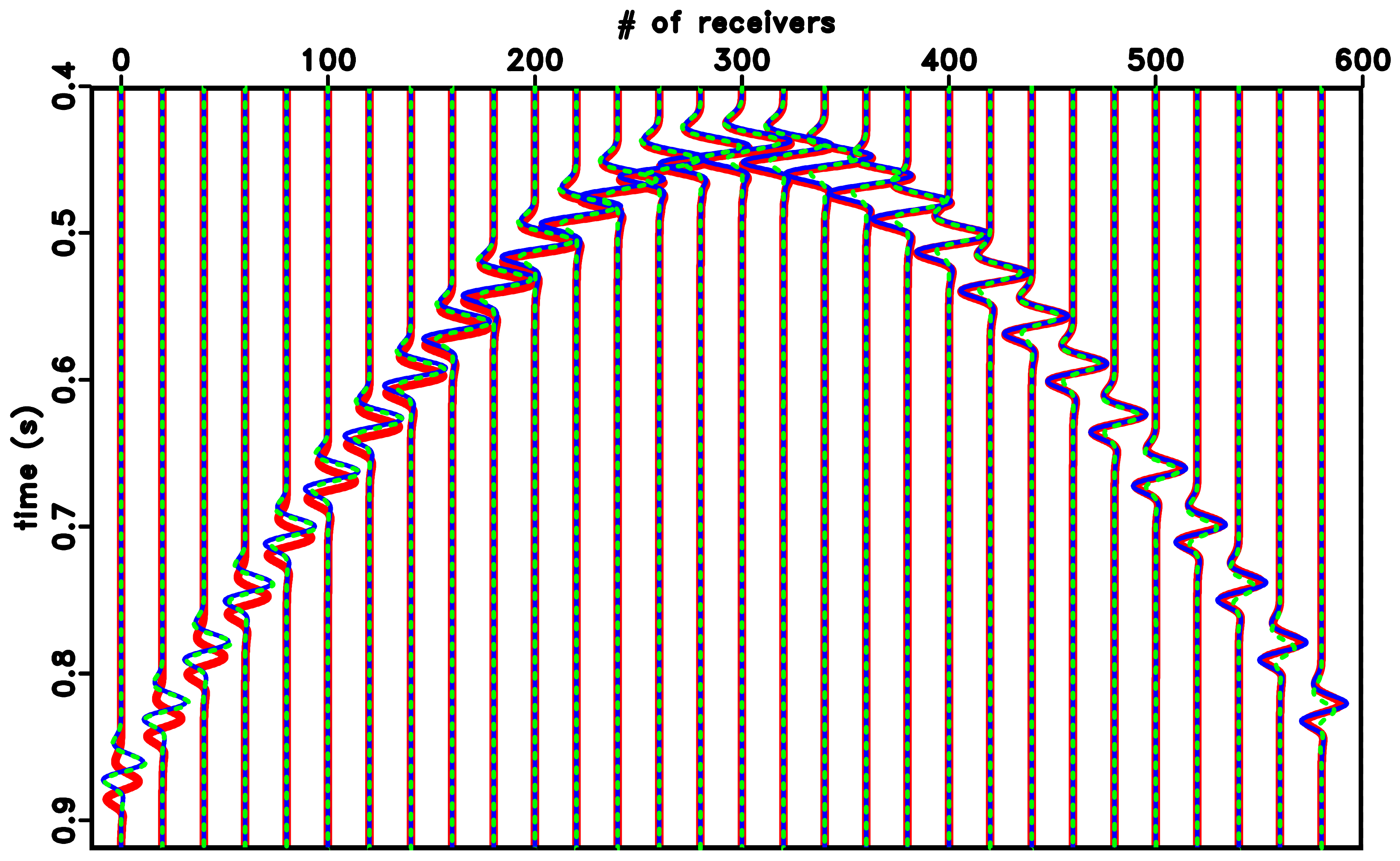}
  \caption{Comparison of the modeled direct waves using the velocity models shown in Figures \ref{fig:velsmooth2} (thin blue lines) and \ref{fig:velsmooth3} (thick red lines), overlain with the direct wave modeled using the true velocity model in Figure \ref{fig:vel1} (dashed green lines). The traces have been multiplied by $exp(2t)$.}
  \label{fig:dirros}
\end{figure}

As described and shown in Sections 2 and 3, we model the direct wave using the background velocity model and start the iterative scheme. To evaluate the differences only in the modeled direct waves using different velocity models (Figures \ref{fig:velsmooth2}, \ref{fig:velsmooth3}), we show in Figure \ref{fig:dirros} the comparison of the modeled direct waves using the velocity models shown in Figures \ref{fig:velsmooth2} (thin blue lines) and \ref{fig:velsmooth3} (thick red lines), overlain with the direct wave modeled using the true velocity model in Figure \ref{fig:vel1} (dashed green lines). For the receivers between 100 and 500 in Figure \ref{fig:dirros}, the modeled direct waves are almost identical for the less smoothed velocity model (thin blue lines) and constant velocity model (thick red line) with the true velocity model (dashed green line). This high similarity in the modeled direct waves also produces the high-accuracy CCs (Figures \ref{fig:cclesssmooth}, \ref{fig:cccons}).

\begin{figure}
  \centering
  \includegraphics[width=0.9\textwidth]{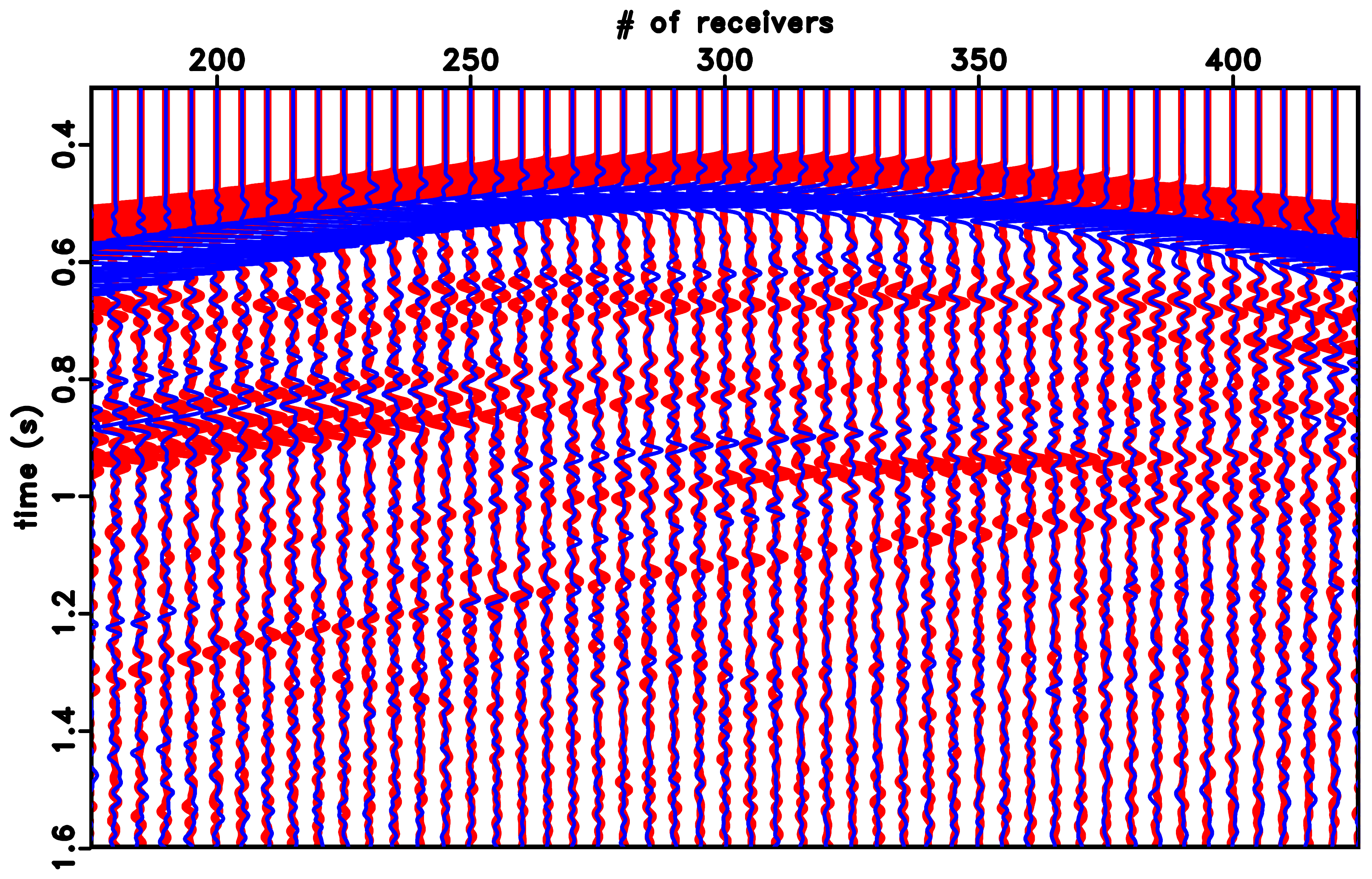}
  \caption{The retrieved Green's function using the constant background velocity model with 10\% error (blue lines) and the modeled Green's function (red lines). The traces have been multiplied by $exp(2t)$ to emphasize the match for the later times.}
  \label{fig:wiggle-shift}
\end{figure}

We further test the sensitivity of the Marchenko method to the velocity model by adding a 10\% error to the velocity model shown in Figure \ref{fig:velsmooth3} and assume the constant velocity as 1.5 km/s. We show the retrieved Green's function using the constant background velocity model with 10\% error (thin blue lines) and the modeled Green's function (thick red lines) superimposed in Figure \ref{fig:wiggle-shift} after multiplying traces by $exp(2t)$. The mismatch in time, amplitudes, and phase in Figure \ref{fig:wiggle-shift} indicate that the constant value of the velocity should be calculated using the average slowness between the surface and the depth of the focal point, and erroneous constant velocity models will not retrieve the accurate Green's function. 

\begin{figure}
  \centering
  \includegraphics[width=0.6\textwidth]{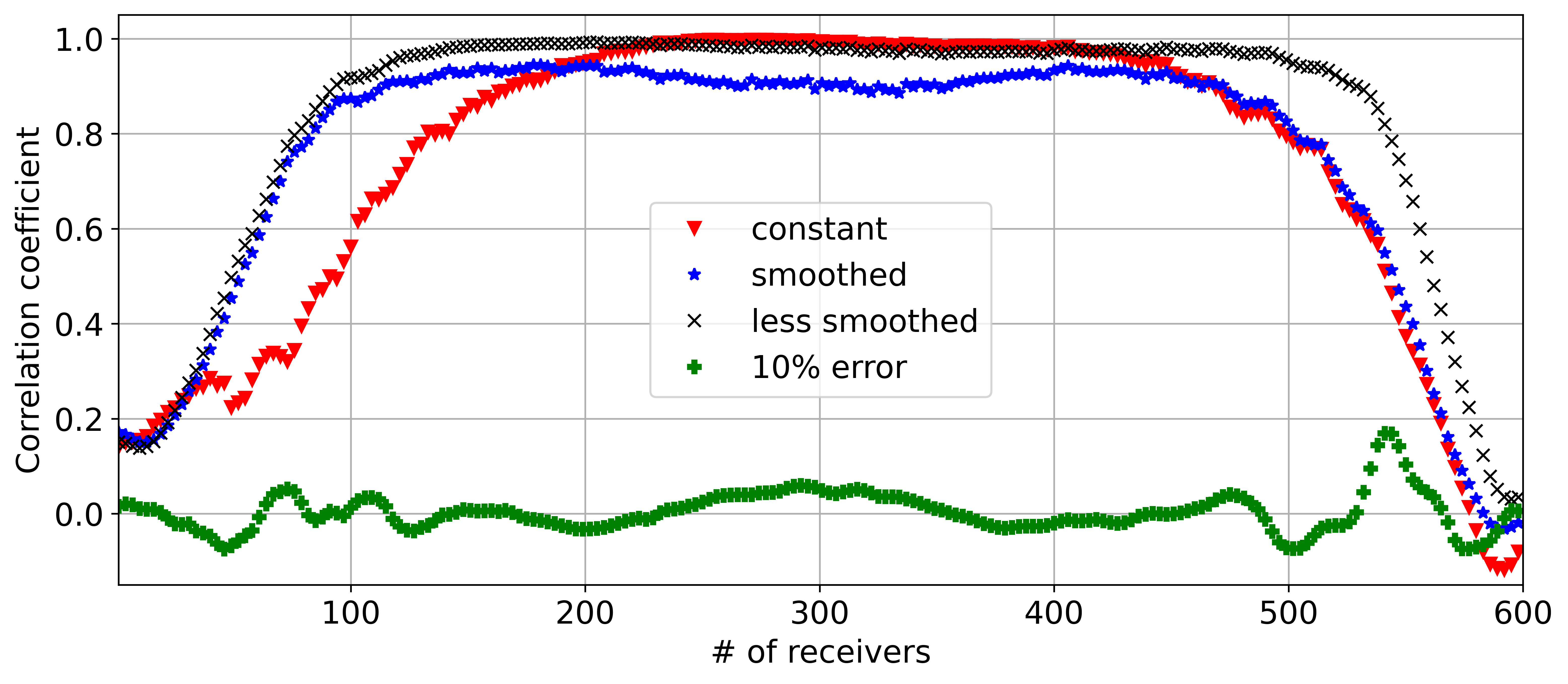}
     \caption{Trace-by-trace calculated between the numerically modeled Green's function (Figures \ref{fig:gf}, \ref{fig:gf-less}, and \ref{fig:gf-cons}) and the retrieved Green's functions using the velocity models from Figures \ref{fig:velsmooth1}, \ref{fig:velsmooth2}, \ref{fig:velsmooth3}, and the velocity model in Figure \ref{fig:velsmooth3} with 10\% error.}
  \label{fig:ccfull}
\end{figure}

Lastly, we present the CCs in Figure \ref{fig:ccfull} between the numerically modeled Green's function and the retrieved Green's functions using the velocity models from Figures \ref{fig:velsmooth1}, \ref{fig:velsmooth2}, \ref{fig:velsmooth3}, and the velocity model in Figure \ref{fig:velsmooth3} with 10\% error using the blue star markers, the grey cross markers, the red triangle markers, and the green plus markers, respectively. We see in Figure \ref{fig:ccfull} that the similarity in the modeled direct wave for the Marchenko focusing creates a high accuracy in the retrieved Green's functions. The star, the cross, and the triangle markers show CCs around 0.9; however, the plus marker shows a CC around 0. Therefore, we conclude that the Green's function retrieval using the Marchenko equation successfully retrieves the Green's functions as long as the correct average slowness between the surface and the depth of the focal point is known.

\begin{figure}
  \centering
  \subfigure[]{\includegraphics[width=0.6\textwidth]{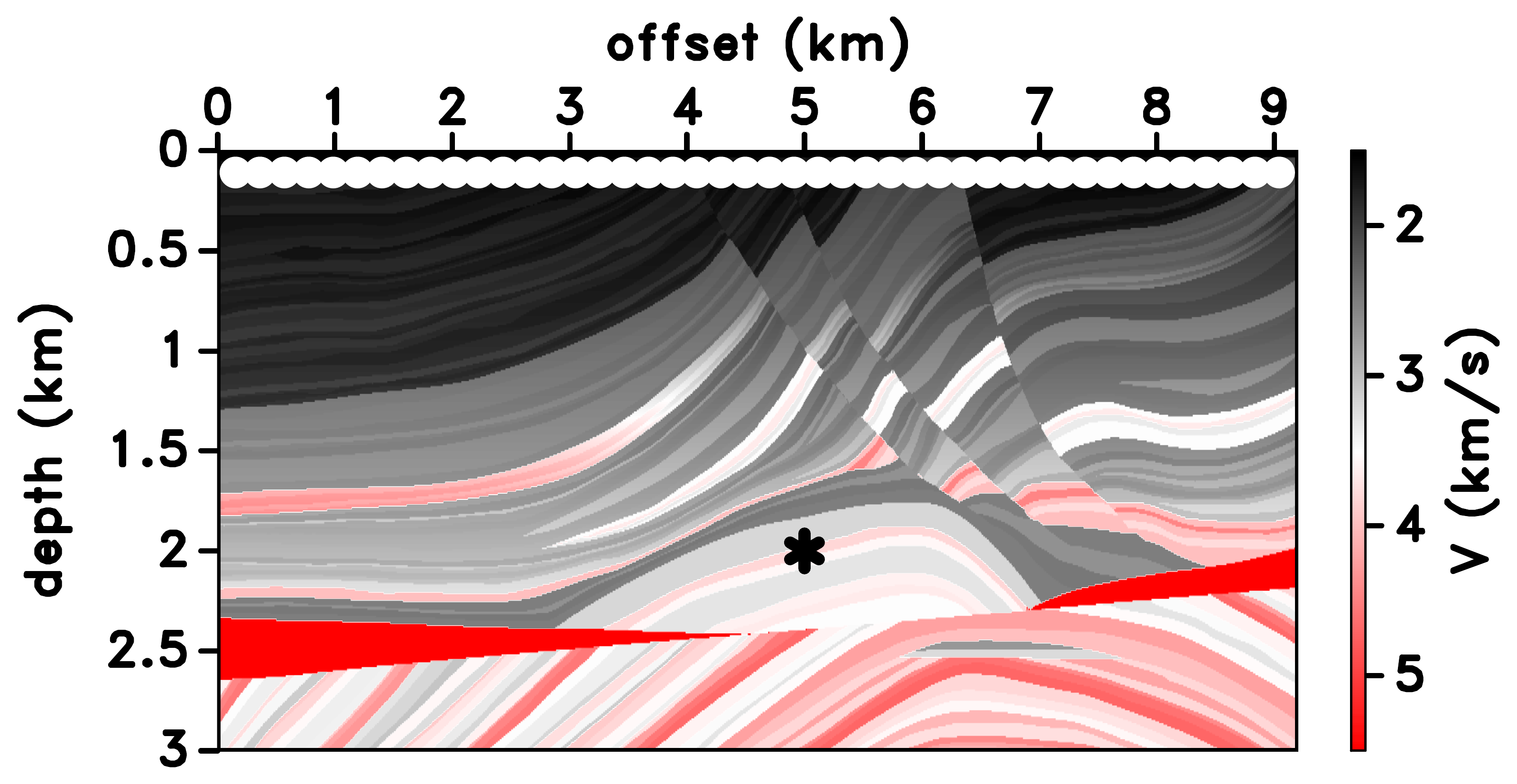}\label{fig:mvel1}}
  \subfigure[]{\includegraphics[width=0.6\textwidth]{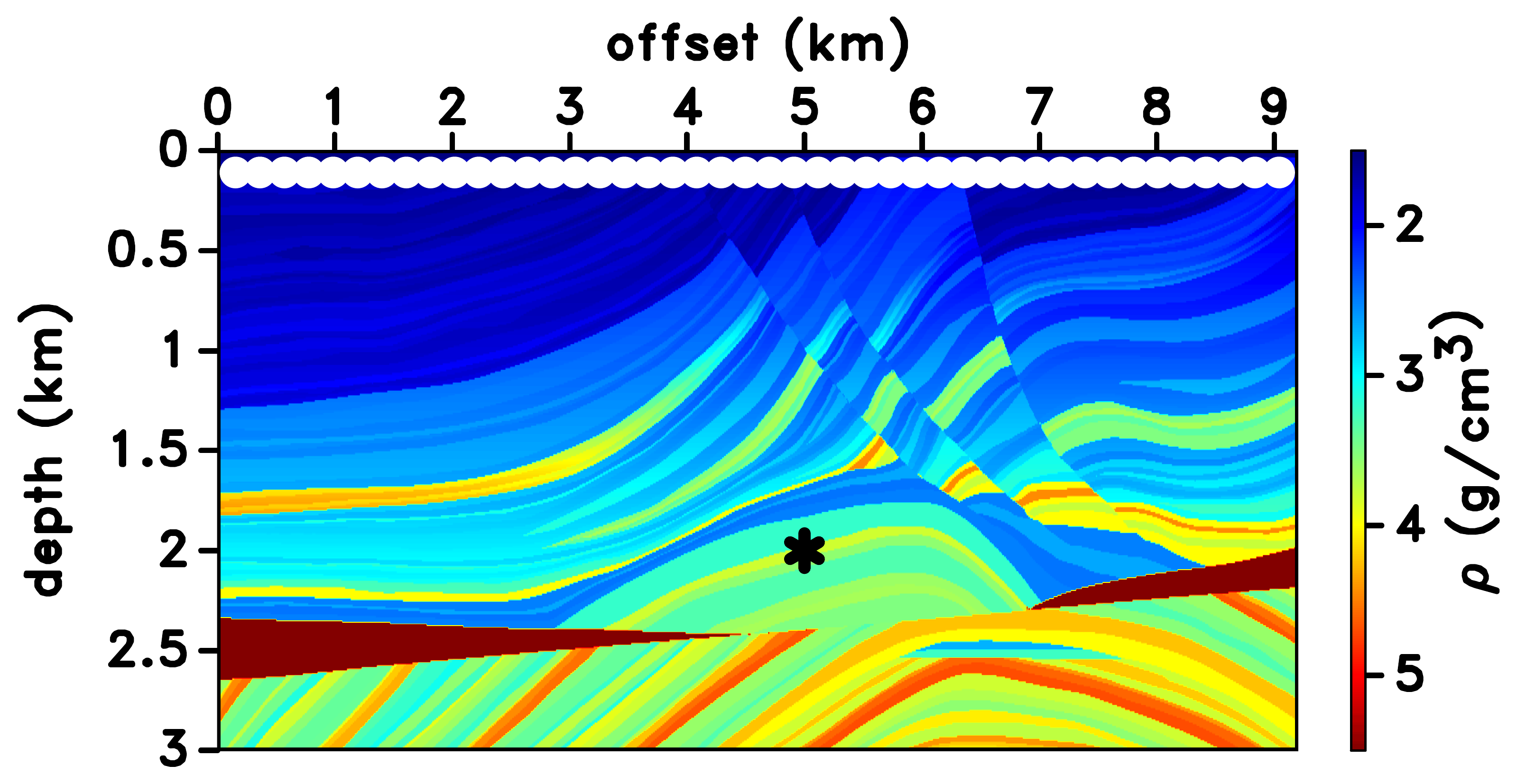}\label{fig:mden1}}
  \caption{(a) Marmousi velocity model and (b) density model of the synthetic example. The black asterisk shows the virtual source location and the white dots at the top indicate every 30th source/receiver location.}
  \label{fig:mmodels}
\end{figure}
\section{Refracted waves}

In this section, we investigate the refracted wave presence in the retrieved Green's function using the Marchenko focusing. To model refracted waves in the Green's functions, we use the Marmousi velocity model \cite[]{marmousi} for the numerical experiments in this section. Figure \ref{fig:mvel1} shows the Marmousi velocity model and Figure \ref{fig:mden1} shows the density model of our experiment. The white dots in Figure \ref{fig:mmodels} represent every 30th receiver location at the surface and the black asterisk denotes the virtual source location for which the Green's function will be retrieved. 

\begin{figure}
  \centering
  \includegraphics[width=0.6\textwidth]{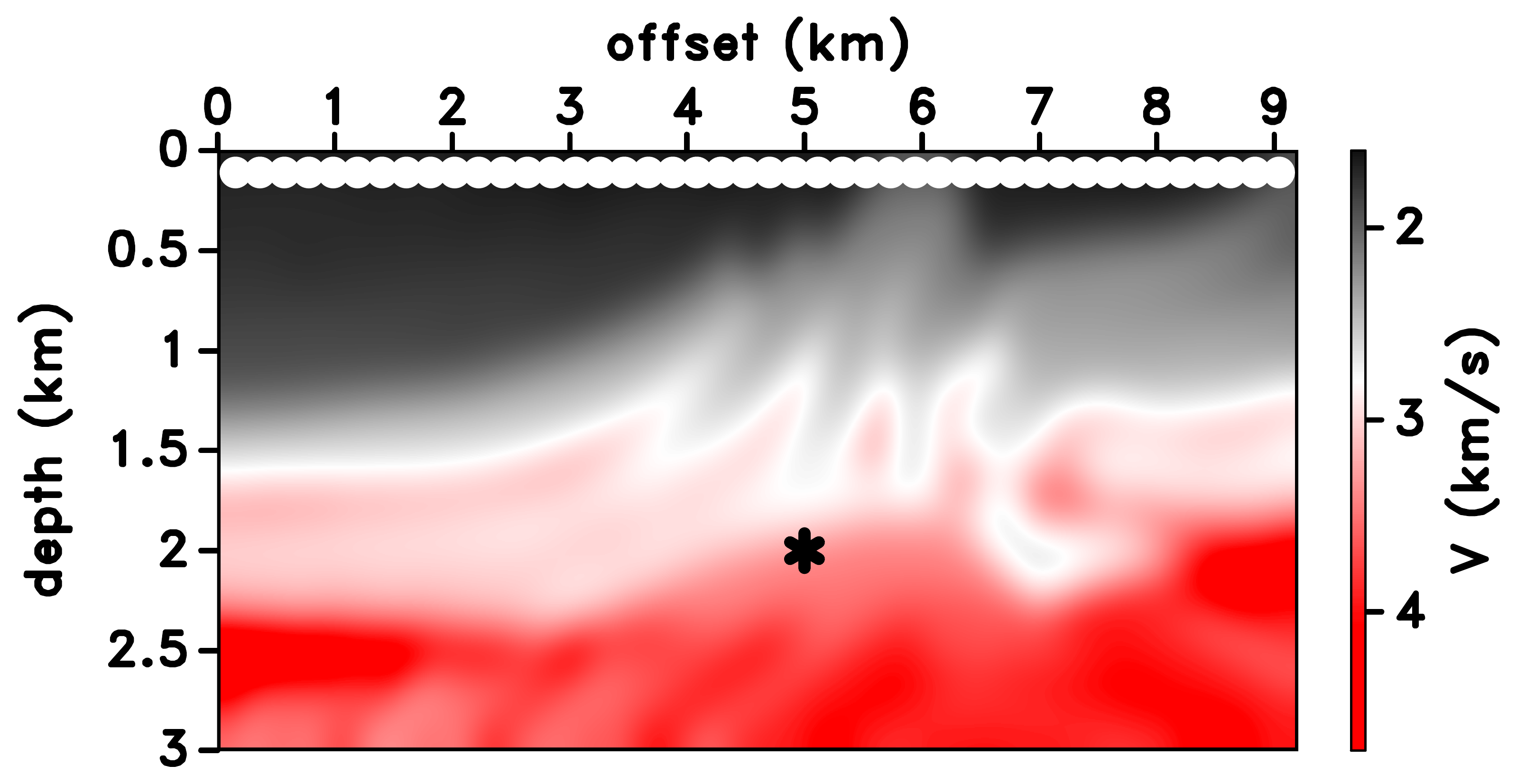}
  \caption{Smoothed version of the Marmousi velocity model used for the iterative algorithm. The black asterisk shows the virtual source location and the white dots at the top indicate every 30th source/receiver location.}
  \label{fig:msmooth}
\end{figure}

\begin{figure}
  \centering
  \subfigure[]{\includegraphics[width=0.5\textwidth]{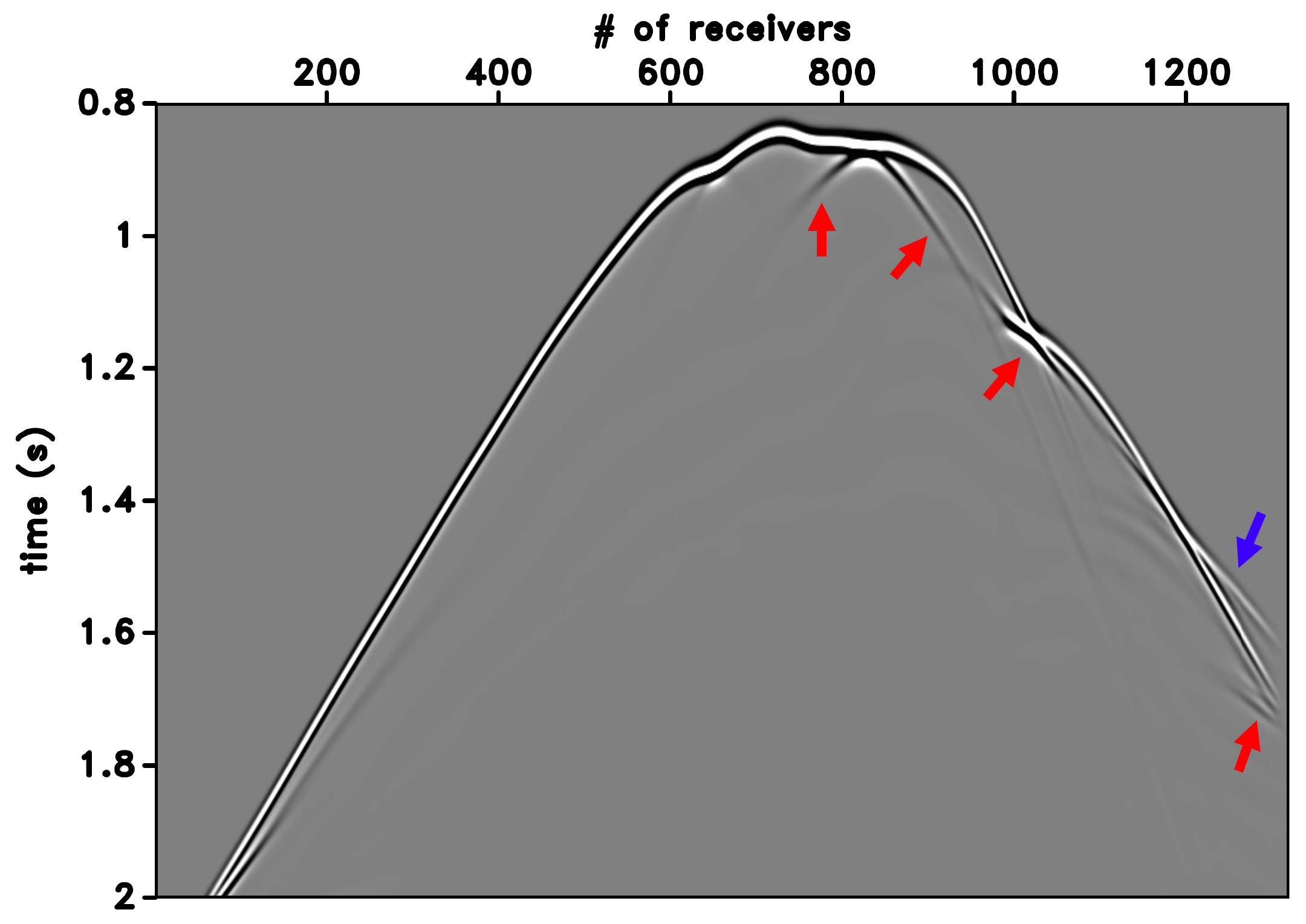}\label{fig:mdir}}
  \subfigure[]{\includegraphics[width=0.5\textwidth]{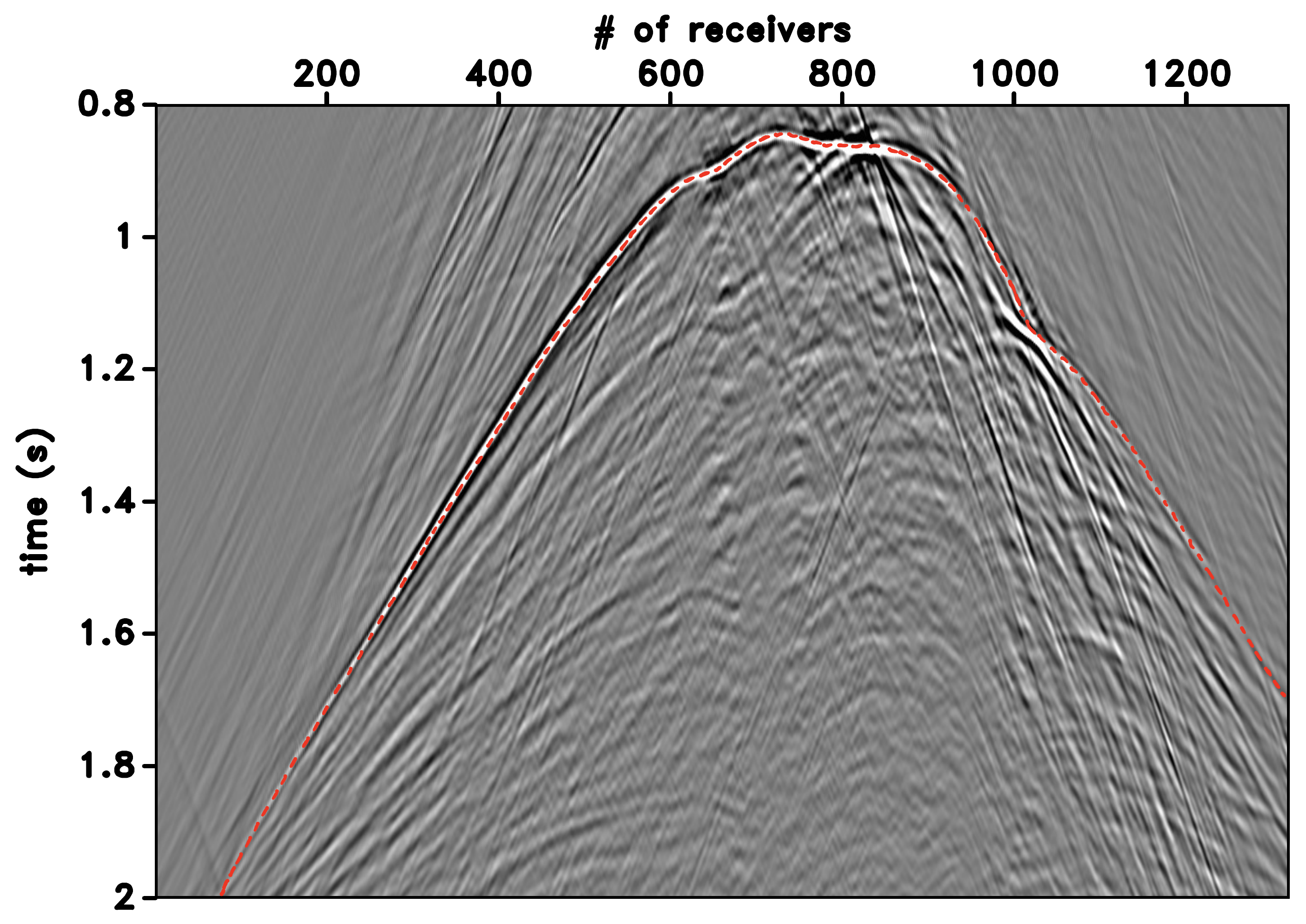}\label{fig:mretr}}
  \subfigure[]{\includegraphics[width=0.5\textwidth]{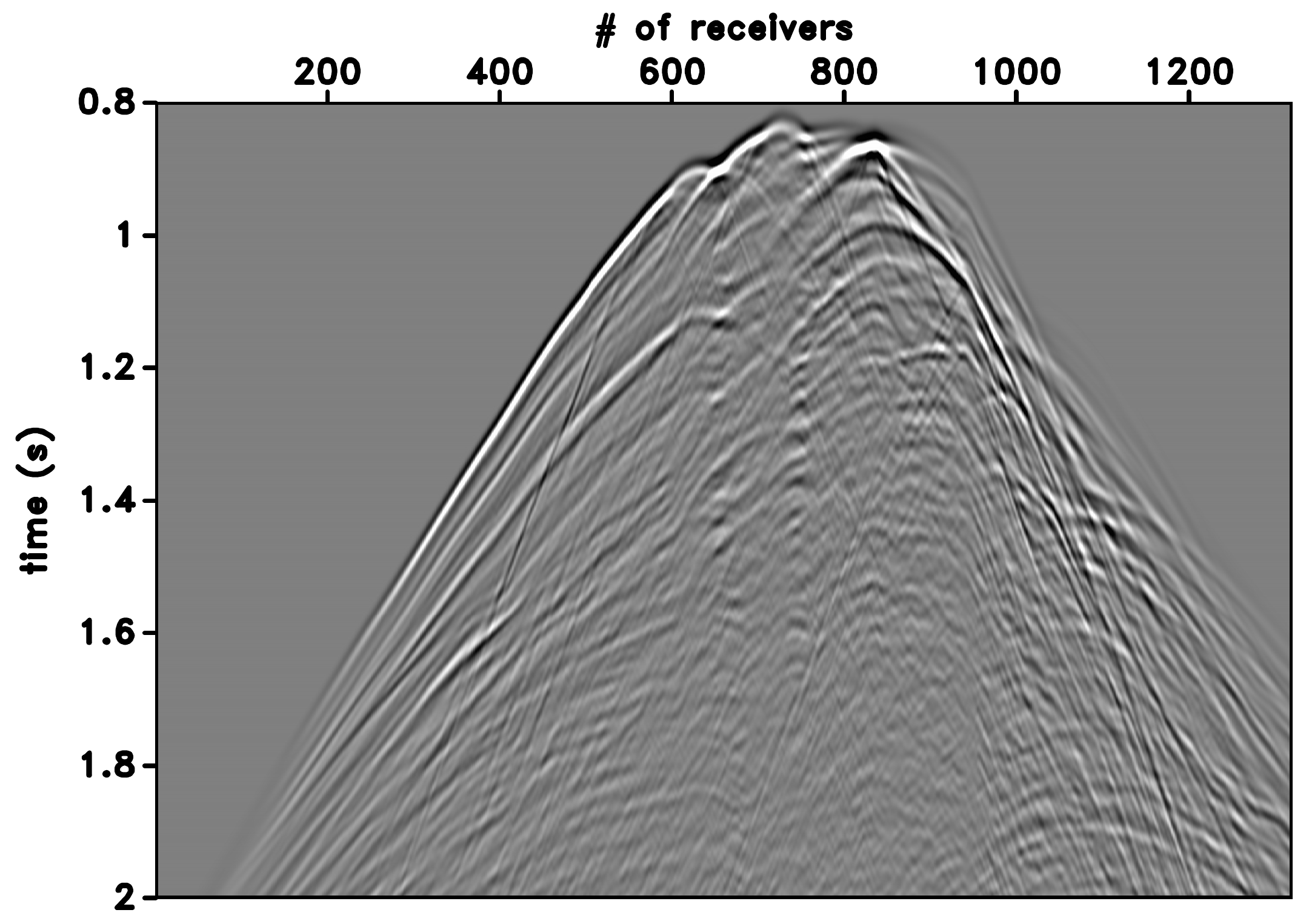}\label{fig:mgf}}
  \caption{(a) Modeled direct wave using the smoothed version of the Marmousi velocity model given in Figure \ref{fig:msmooth}. The red arrows show some of the triplicated arrivals and the blue arrow shows the refracted wave. (b) Retrieved Green's function using the Marchenko focusing using the smoothed version of the Marmousi velocity model. The red dashed curve shows the arrival of the direct wave (including some triplicated waves), approximately. (c) Numerically modeled Green's function.}
  \label{fig:marms}
\end{figure}

To start the iterative algorithm, we model the direct wave using the smoothed background velocity model shown in Figure \ref{fig:msmooth}, and the modeled direct wave is shown in Figure \ref{fig:mdir}. The red arrows in Figure \ref{fig:mdir} point to some of the triplicated arrivals and the blue arrow points out the refracted wave. After following the iterative scheme, we show the retrieved Green's function after four iterations in Figure \ref{fig:mretr}, and the numerically modeled Green's function for the virtual source location in Figure \ref{fig:mgf}. The red dashed curve in Figure \ref{fig:mretr} indicates the arrival of the direct wave (including some triplicated waves), and the waves before the red dashed curve in Figure \ref{fig:mretr} can be removed by applying a muting function. The main difference in Figure \ref{fig:marms} between the retrieved and the modeled Green's functions (Figures \ref{fig:mretr} and \ref{fig:mgf}, respectively) occurs between the receivers 800 and 1320 around the arrival time of the direct wave (around the red dashed curve in Figure \ref{fig:mretr}). The numerically modeled Green's function (Figure \ref{fig:mgf}) contains refracted and horizontally propagating events recorded between the receivers 800 and 1320 before the arrival time of the direct wave (also indicated with the red dashed curve in Figure \ref{fig:mretr}); however, the refracted and horizontally propagating events are not present before the arrival time of the direct wave in the retrieved Green's function (Figure \ref{fig:mretr}). 

\begin{figure}
  \centering
  \includegraphics[width=0.6\textwidth]{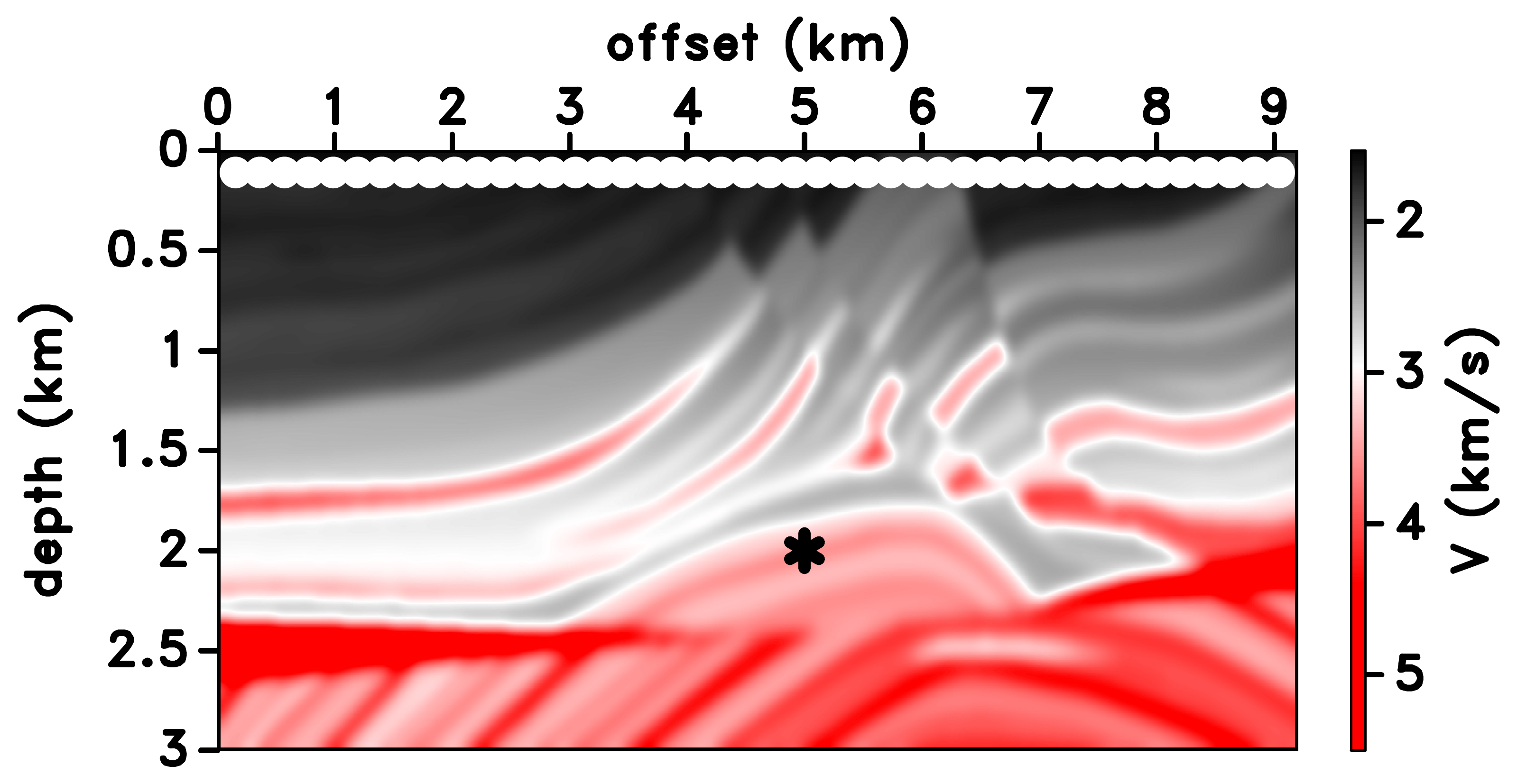}
  \caption{Less smoothed version of the Marmousi velocity model used for the iterative algorithm. The black asterisk shows the virtual source location and the white dots at the top indicate every 30th source/receiver location.}
  \label{fig:msmooth-less}
\end{figure}

\begin{figure}
  \centering
  \subfigure[]{\includegraphics[width=0.5\textwidth]{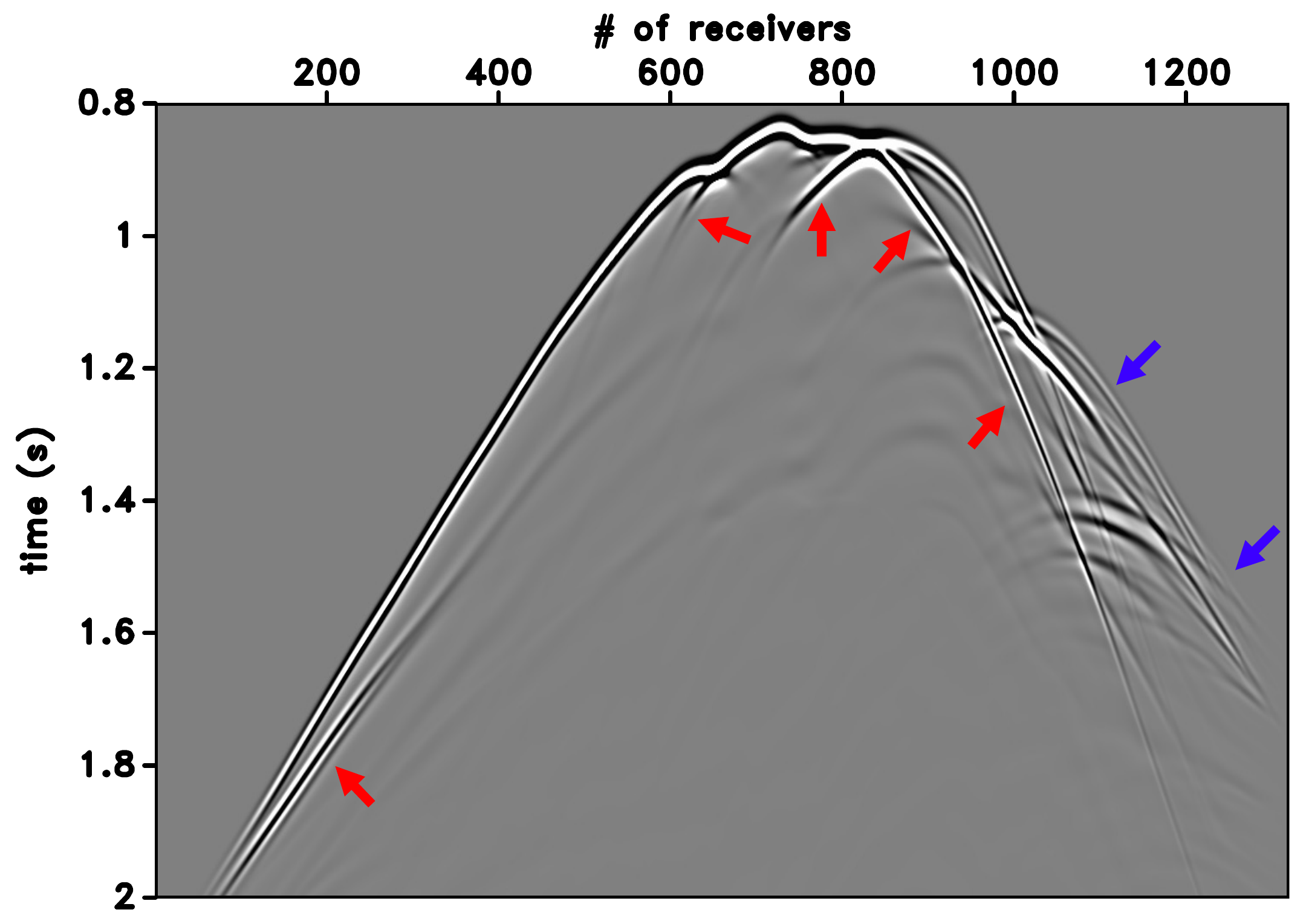}\label{fig:mdir-less}}
  \subfigure[]{\includegraphics[width=0.5\textwidth]{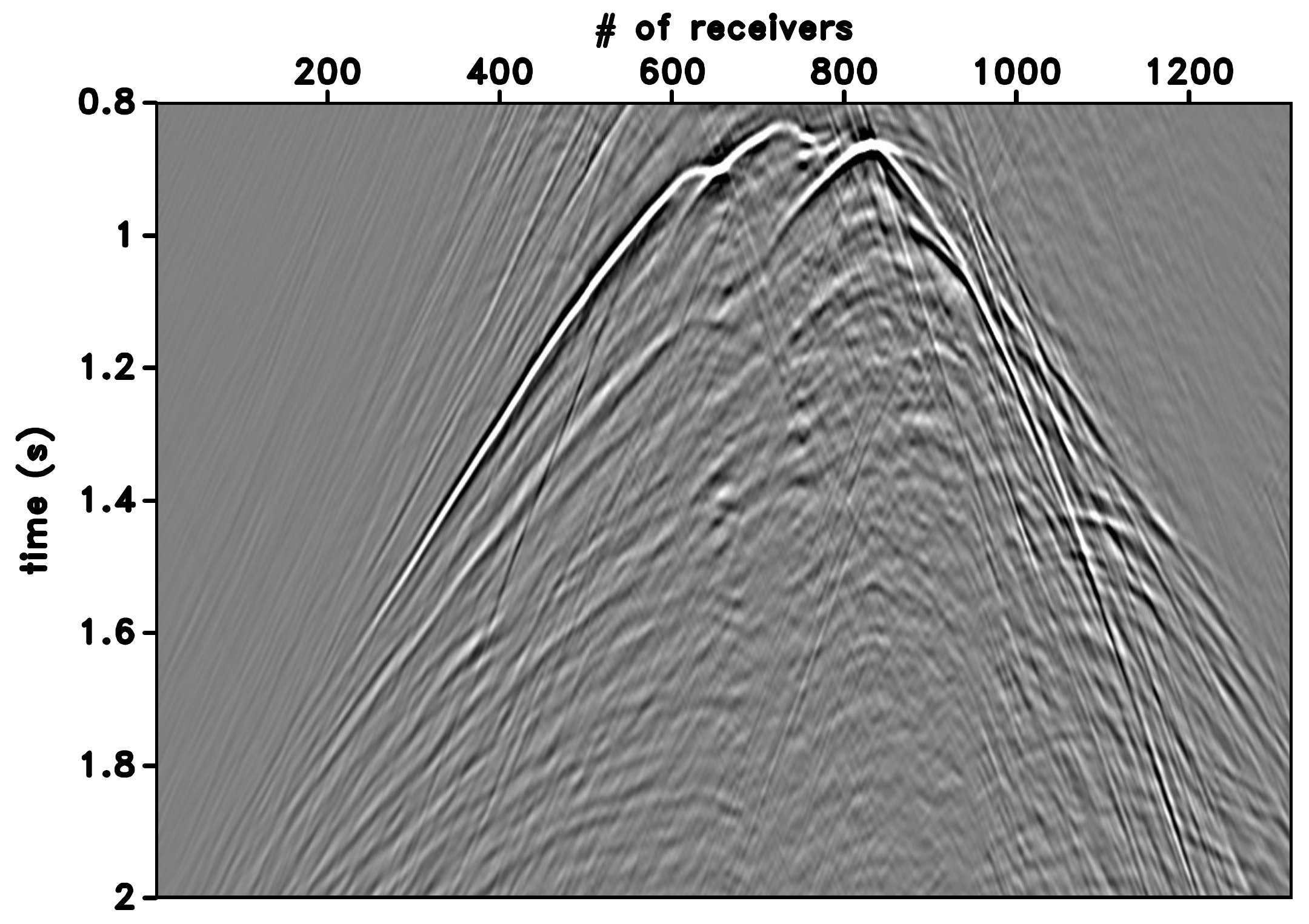}\label{fig:mretr-less}}
  \subfigure[]{\includegraphics[width=0.5\textwidth]{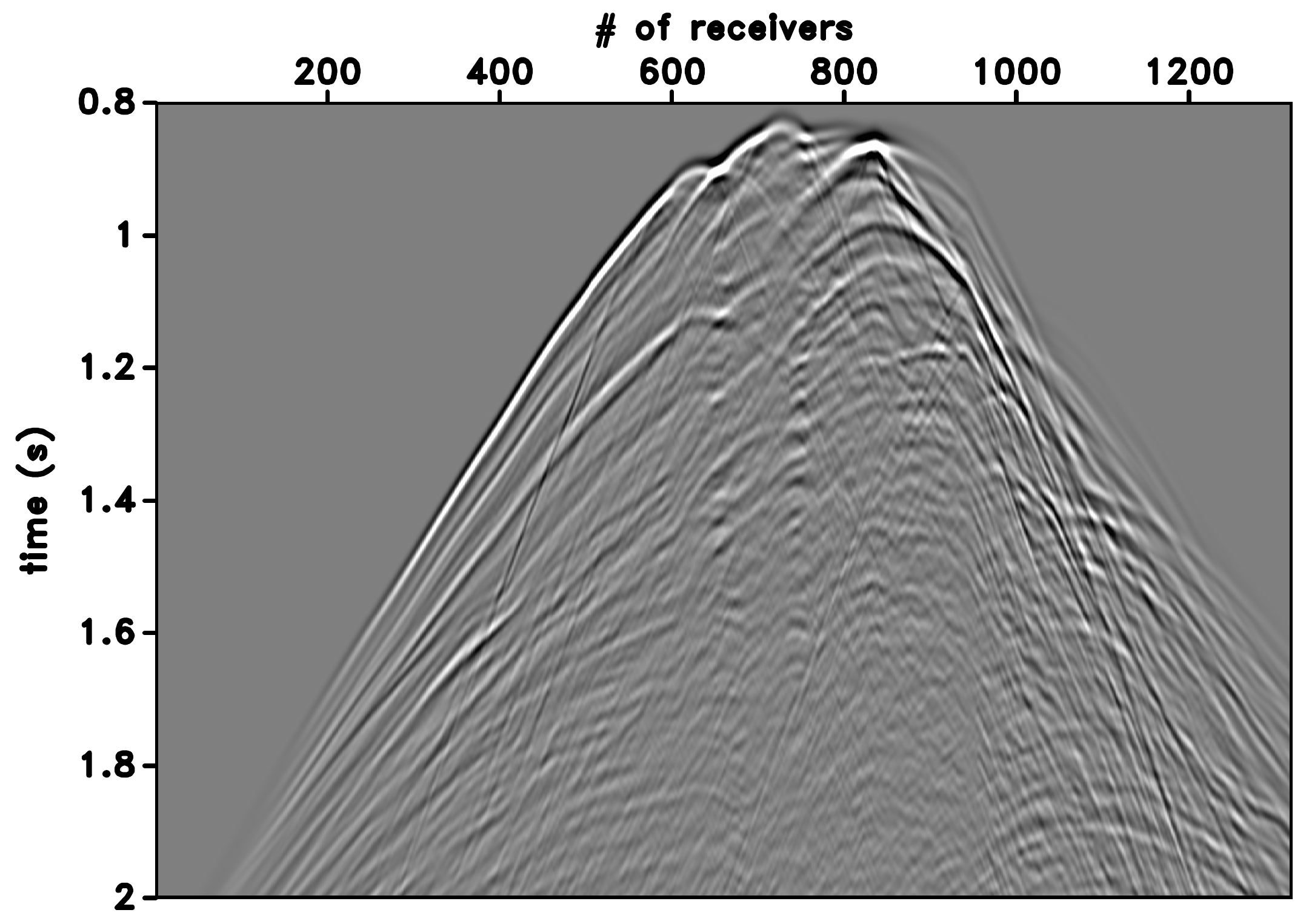}\label{fig:mgf-less}}
  \caption{(a) Modeled direct wave using the less smoothed version of the Marmousi velocity model given in Figure \ref{fig:msmooth-less}. The red arrows show some of the triplicated arrivals and the blue arrows show the refracted wave. (b) Retrieved Green's function using the Marchenko focusing using the less smoothed version of the Marmousi velocity model. (c) Numerically modeled Green's function (which is the same wavefield as Figure \ref{fig:mgf}).}
  \label{fig:marms-less}
\end{figure}

To further investigate the presence of the refracted wave in the Marchenko focusing, we use a less smoothed version of the Marmousi velocity model than the one shown in Figure \ref{fig:msmooth} which presents more detailed subsurface information. Figure \ref{fig:msmooth-less} shows the less smoothed Marmousi model and the white dots show every 30th receiver location and the black asterisk denotes the virtual source location. The modeled direct wave using the less smoothed Marmousi model is shown in Figure \ref{fig:mdir-less}, and the red arrows indicate some of the triplicated arrivals, and the blue arrows indicate the refracted wave. Figure \ref{fig:mretr-less} shows the retrieved Green's function by using the direct wave in Figure \ref{fig:mdir-less}, and Figure \ref{fig:mgf-less} shows the numerically modeled Green's function. This time, between the receivers 800 and 1320 and around the direct arrival times, the retrieved Green's function and the numerically modeled Green's function match very well. The refracted wave information in the modeled Green's function is also present in the retrieved Green's function. The less smoothed version of the background velocity model used for the iterative algorithm enables the refracted waves to appear in the retrieved Green's function.

If we compare the modeled direct waves in Figures \ref{fig:mdir} and \ref{fig:mdir-less}, the modeled direct wave in Figure \ref{fig:mdir} does not include most of the refracted waves (events shown with blue arrows in Figure \ref{fig:mdir-less}). However, the modeled direct wave in Figure \ref{fig:mdir-less} includes the refracted waves shown with the blue arrows. Figures \ref{fig:mretr} and \ref{fig:mretr-less} show that the refracted waves modeled using the smooth velocity model are mapped directly in the retrieved Green's functions. In other words, if the refracted waves are modeled using the background velocity model, those events are also present in the retrieved Green's function. But, if the refracted waves are not present in the modeled direct wave, they are not present in the retrieved Green's function. We conclude that the presence of refracted waves only depends on the background velocity model in the Marchenko focusing and it is not a result of the iterations of the Marchenko focusing algorithm.

\section{Conclusions}
We present the Green's function retrieval using Marchenko focusing and investigate the background velocity model dependence of the Marchenko focusing. We compare the retrieved Green's functions for three different background velocity models used for modeling the direct wave. We show that the Marchenko focusing algorithm can retrieve the Green's function with high accuracy. We also investigate the presence of the refracted waves in the retrieved Green's function using the Marmousi velocity model. We show that the refracted waves are incorporated in the retrieved Green's function by the background velocity model used to model the direct wave for the iterative algorithm, and the Marchenko algorithm does not produce the refracted waves.
\section{Acknowledgements}

We thank the editor Michal Malinowski, assistant editor Fern Storey, reviewer Ole Edvard Aaker and the anonymous reviewer for their constructive reviews. This work is supported by the Consortium Project on Seismic Inverse Methods for Complex Structures at the Colorado School of Mines. The numerical examples in this paper were generated using the Madagascar software package \cite[]{madagascar}. The research of K. Wapenaar has received funding from the European Research Council (grant no. 742703).

\section{Data and materials availability}
Data associated with this research are available and can be obtained by contacting the corresponding author.





\newpage

\bibliographystyle{seg}  
\bibliography{example}

\end{document}